\documentclass[aps,pre,showpacs,groupedaddress,floatfix]{revtex4}
\usepackage{amsmath}
\usepackage{amssymb}
\usepackage{epsfig}
\usepackage{amscd}
\usepackage{color}
\usepackage{comment}
\usepackage{ulem}
\usepackage{bm}
\usepackage{pdfsync}

\newcommand{\beq}{\begin{equation}}
\newcommand{\eeq}{\end{equation}}
\newcommand{\bea}{\begin{eqnarray}}
\newcommand{\eea}{\end{eqnarray}}

\newcommand{\flabel}[1]{\label{f:#1}}
\newcommand{\elabel}[1]{\label{e:#1}}

\newcommand{\eq}[1]{Eq.~(\ref{e:#1})}

\newcommand{\Psitilde}{\tilde{\Psi}}

\newcommand{\eqtwo}[2]{Eqs~(\ref{e:#1}) and~(\ref{e:#2})}
\newcommand{\fig}[1]{Fig.~\ref{f:#1}}

\newcommand{\VEC}[1]{\mathbf{#1}}
\newcommand{\Vol}{\Omega}

\newcommand{\kvec}{\VEC{k}}

\newcommand{\nvec}{\VEC{n}}

\newcommand{\qvec}{\VEC{q}}
\newcommand{\rvec}{\VEC{r}}

\newcommand{\Rvec}{\VEC{R}}

\newcommand{\etavec}{\boldsymbol{\eta}}

\begin{document}

\title{ Theory of Finite Size Effects for Electronic Quantum Monte Carlo Calculations of Liquids and Solids}

\author{Markus Holzmann$^{123}$}
\author{Raymond C. Clay  III$^4$}
\author{Miguel A. Morales$^5$}
\author{Norm M. Tubman$^4$}
\author{David M. Ceperley$^4$}
\author{Carlo Pierleoni$^6$}
\affiliation{$^1$LPTMC, UMR 7600 of CNRS, Universit\'e Pierre et Marie Curie, F-75005 Paris, France\\
$^2$LPMMC, UMR 5493 of CNRS, UniversitŽ Grenoble Alpes,  F-38042
Grenoble, France\\
$^3$Institut Laue-Langevin, BP 156, F-38042 Grenoble Cedex 9, France \\
$^4$University of Illinois Urbana-Champaign,Urbana, Illinois 61801, USA \\
$^5$Lawrence Livermore National Laboratory, Livermore, California 94550, USA\\
$^6$Department of Physical and Chemical Sciences, University of L'Aquila, Via Vetoio 10, I-67010 L'Aquila, Italy}
\date{\today}

\begin{abstract}
Concentrating on  zero temperature Quantum Monte Carlo calculations of
electronic systems, we give a general description of the theory of finite size extrapolations of energies to the thermodynamic limit based on one and two-body correlation functions. 
We introduce new effective procedures, such as using the potential and wavefunction split-up into long and short range functions to simplify the method and we discuss how to treat backflow wavefunctions.   Then we explicitly test the accuracy of our method to correct finite size errors on example  hydrogen and helium many-body systems and show that the finite size bias can be drastically reduced for even small systems. 
\end{abstract}

\maketitle

Quantum Monte Carlo (QMC) methods allow us to calculate the energy per particle $E_N$ of a finite system containing
$N$ particles, with  $N \lesssim 10^{3}$ for almost all simulations of electronic systems \cite{RMP,Lubos}. 
However, for extended systems, we are often interested in scaling to
the thermodynamic limit, $E_\infty$; this scaling is one of the major source of bias in Quantum Monte Carlo calculations of electronic structure. In practice, extrapolation is often performed
numerically by assuming simple functional forms for $E_N$ as a function of $1/N$, often inspired by
results of approximate theories, such as Kohn-Sham DFT \cite{Rajagopal,MPC,Shiwei}   or from the behavior of approximate many-body calculations, e.g.  from RPA calculations \cite{fse_rpa}. 
These heuristic extrapolations can be dangerous and introduce a possible systematic bias, as the  exact ground state energy, as well as other properties, are in general not a simple analytic function of $1/N$. In fact, the scaling function will depend on the electronic state, for example, it will be different in a metal and an insulator, and can depend on the form of the trial wave function underlying the QMC calculation.
In addition, within variational approaches, the amount that the variational energy is above the exact energy may depend on the system size because of the values of the variational parameters. This   introduces a
further source of error in a purely numerical extrapolation.
Projection methods can reduce this bias, since they are closer to the true ground state energy, but in practice it can be a difficult problem to ensure a uniform convergence concerning projection time or population size with respect to the system size \cite{Saverio-Massimo}.

In this paper we present a general theory for understanding the finite size bias of QMC calculations. Although we concentrate on
electronic systems where finite size effects represent one of the major limitations, our approach applies equally well
to other quantum systems with different interactions and dimensionality, including bosonic ones.  As we will show, the leading order size effects can be understood by looking at the analytical structure
of the trial wavefunction \cite{fse,fse_ke} which is -- at least partially -- determined by singularities of the Hamiltonian and/or the boundary conditions \cite{Gaskell,ceperley78,bf}. Different types of wavefunction will, in general, have different types of size effects. In particular, we show that
backflow wavefunctions give rise to kinetic energy corrections which have not been considered previously.

Size effects  depend on the observable we are interested in. In this paper we focus on one of the most fundamental quantities, the total energy \cite{fse,drummond,luke,azadi}. However, the ideas can be generalized 
to determine the finite size effects of different observables such as the momentum distribution\cite{momk2d,momk3d,Na,Renyi}. 

The paper is organized as follows. Section I is a general introduction to finite size effects of quantum systems. Focusing on electronic
calculations 
we systematically discuss the origins of size
effects for the kinetic and potential energies.  In Section II we propose robust procedures to estimate the finite size bias in electronic structure calculations. In Section III, we apply our method to QMC calculations of b.c.c. solid hydrogen with different underlying
trial wavefunctions and calculations of liquid hydrogen and hydrogen-helium mixtures. Conclusions are presented in Section IV. Technical aspects concerning the long-range character of the Jastrow potential, the backflow corrections and details of the split-up of the long and short-range part of the potential energy via the Ewald potential are given in the Appendices. 

\section{Introduction to Finite Size Estimation}
Consider $N_a$ particles of species $a=1,\dots M$ with mass $m_a$ described by the Hamiltonian $H=T+V$
where
\bea
T &=&- \sum_{a=1}^M\sum_{i=1}^{N_a} \frac{\hbar^2}{2m_a} \nabla_{i_a}^2  \\
V &=& \sum_a \sum_{i_a < j_a} v_{aa}(\rvec_{i_a}-\rvec_{j_a}) + \sum_{a<b} \sum_{i_a,j_b} v_{ab}(\rvec_{i_a}-\rvec_{j_b})
\eea
are kinetic and potential energy operators. Here, $\rvec_{i_a}$ are the coordinates of species $a$, 
$v_{aa}$ ($v_{ab}$) are the intra (inter) species potentials.
The energy per particle is then
\beq
E_{N}= \langle H \rangle_0 /N
\eeq
where $\langle \dots \rangle_0$ denotes its expectation value in the state
$\Psi_0(\Rvec)$.   Here, $\Rvec$ indicates the coordinates of all particles and we take the normalization $N$ to be the total number of electrons, $N=N_e$, but it must be proportional to the volume, $\Vol$; one takes the thermodynamic limit so that the density is fixed in the limit of $N \rightarrow \infty$.

Since the Hamiltonian only involves single and two-body potentials,
we can express the total energy in terms of the reduced single particle and two-particle density matrices of $\Psi_0$:
$\rho^{(1)}_a(\rvec;\rvec')$ and $\rho^{(2)}_{ab}(\rvec_1,\rvec_2;\rvec_1',\rvec_2')$ (in the latter we will only need the diagonal components $\rvec_1=\rvec_1'$ and $\rvec_2=\rvec_2'$).
We write the reduced density matrices in Fourier space in terms of the momentum distribution, $n_\kvec^a$, and the structure factor $S_{ab}(\kvec)$ 
\bea
n_\kvec^a&=& \frac{1}{\Vol} \int d\rvec \int d\rvec' e^{i \kvec \cdot (\rvec-\rvec')} \rho^{(1)}_a(\rvec;\rvec')
\\
S_{ab}(\kvec) &=& \delta_{ab}+ \frac{1}{N} \int d\rvec \int d \rvec' e^{i \kvec   \cdot (\rvec-\rvec')}  \rho^{(2)}_{ab}(\rvec,\rvec';\rvec,\rvec')
= \frac{1}{N} \langle \rho_\kvec^a \rho_{-\kvec}^b \rangle
\eea 
where $\rho_\kvec^a=\sum_{i_a} \exp[i \kvec \cdot \rvec_{i_a}]$ are collective density fluctuations.
Here, and in the following, we use the convention
\bea
v(\rvec)&=&\frac{1}{\Vol} \sum_\kvec v_\kvec e^{i \kvec \cdot \rvec}
\\
v_\kvec&=& \int d\rvec \, e^{-i \kvec \cdot \rvec} v(\rvec)
\eea
for discrete Fourier transforms of periodic functions inside a volume $\Vol$. The kinetic and potential
energy per particle can then be expressed as
\bea
\label{eq:T_N}
T_N&=& \frac{1}{N} \sum_a \sum_\kvec \frac{\hbar^2 k^2}{2m_a} n_\kvec^a \\
V_N &=& \frac{1}{2 \Vol} \sum_{a,b} \sum_{\kvec} v_\kvec^{ab} \left[S_{ab}(\kvec)-N_a/N\delta_{ab} \right].
\label{eq:V_N}
\eea

These expressions are our basis for understanding the size effects in periodic boundary conditions.
To simplify the notation, we restrict to a single component system and consider the
the
momentum distribution $n_k^N$ and the structure factor $S_N(k)$ for a finite system of $N$ particles in a cube of linear extension $L$. 
With periodic boundary conditions, both functions
are given on a discrete grid in Fourier space of spacing $2\pi/L$. 
In the thermodynamic limit these functions attain their limiting forms:
$n_\kvec^N \to n_\kvec^\infty$ and $S_N(\kvec) \to S_\infty(\kvec)$.
Assuming a smooth behavior of  $n_\kvec^N $ and $S_N(\kvec)$ as a function of $\kvec$, 
their interpolation 
$\tilde{n}_\kvec^N$ and $\tilde{S}_N(\kvec)$,  to 
all $\kvec$ values should provide the best estimate of the finite system values to $n_k^\infty$ and $S_\infty(\kvec)$.  Note that special care is warranted in the interpolation near non-analytic values of $\kvec$ such as $\kvec=0$ or at the Fermi surface of a metal. 
Then there are  two different ways finite size errors can arise: 
i) changes in the correlation functions as a function of $N$, differences of $\tilde{n}_\kvec^N$ and $\tilde{S}_N(\kvec)$ from
their values in the thermodynamic limit;
ii) differences resulting from a summation of k-points on a finite mesh in reciprocal space rather than an integration.
Changes (i.e. the first way)  in the correlation functions are expected for system sizes smaller than a characteristic correlation length. In particular, close to phase transitions, the correlation length can get large or even diverge,
and finite size extrapolation methods based on additional scaling assumptions have been developed for these cases in the
field of critical phenomena \cite{Zinn,Binder}.
For fermionic, and in particular electronic matter, important finite size effects remain even in the case the system size
exceeds the characteristic correlation length because of the second reason; this article exclusively deals with methods to eliminate this finite size bias: the errors in the kinetic and potential energy are simply quadrature errors due to the discrete underlying mesh in Fourier space  and can be expressed as
\bea
\Delta T_N &= & T_\infty-T_N=\frac{1}{\rho} \left[ \int \frac{d\kvec}{(2\pi)^d} - \frac{1}{\Vol} \sum_{\kvec} \right]
 \frac{\hbar^2 k^2}{2m} \tilde{n}_{\kvec}^N 
 \\
 \Delta V_N &=&\Delta V_\infty-V_N= \left[  \int \frac{d\kvec}{(2\pi)^d} - \frac{1}{\Vol} \sum_{\kvec} \right] \frac{v_k}{2} \tilde{S}_N(\kvec)
\eea
where $\rho=N/\Vol$ is the density and $d$ is the spatial dimension.

In order to actually apply these formulas, we need a method to interpolate $n_{\kvec}^N$
and $S_N(\kvec)$ from the grid where we have simulation data to the continuum.
For local functions such as the structure factor,
it would seem easy to interpolate $S_N(\kvec)$. 
However, since the momentum distribution is a non-local quantity, this direct procedure fails.
Indeed proper size extrapolation of the momentum distribution  is slightly more involved  \cite{fse_ke,momk2d,momk3d}
as we have to express first the momentum distribution  in terms of local correlation functions which can be interpolated
more safely. As long as one is only interested in the kinetic energy, it is easier to express
the kinetic energy in terms of a different -- purely local -- estimator and discuss size effects using them.

Shell effects usually dominate finite size bias of the kinetic energy; they can be drastically reduced by employing twist averaged boundary conditions \cite{Lin01}.
For electronic systems, 
 the leading order size corrections  of the potential and kinetic energy  \cite{fse} beyond shell effects
are determined by the Coulomb singularity, $v_k=2(d-1)\pi e^2/k^2$ for $k \rightarrow 0$ ($e$ is the electron charge). However, for typical  system sizes,
the leading order expressions may not be accurate enough \cite{drummond}. To go beyond leading order,
one has to understand the behavior of the ground state wavefunction.

In the next section, we give a detailed discussion of size effects of electronic systems and develop a robust numerical procedure
for thermodynamic limit extrapolation  of kinetic and potential energy, independent of the particular system under study and optimal for generic calculations.

\section{Energy size corrections}

In this section we will explicitly discuss the finite size error and correction schemes for electronic energies in the
Born-Oppenheimer approximation where the ions only act as a static external potential. 
Let us assume that  for a given ionic configuration  the electronic ground state can be described with the Slater-Jastrow form 
\beq
\Psi_T = D e^{-U}
\eeq
where $D$ ensures the antisymmetry of fermions (usually by a Slater determinant) and $U$ is a many-body symmetric function (for electrons).
Having in mind that the electrons feel an 
external potential created by classical ions, we can  simplify the notation to that of a one-component system.
We will further assume a spin-polarized system, the extension to spin-unpolarized system based on two determinants is straightforward.

The corresponding ``local energy'' is:
\beq
E_L=\frac{H\Psi_T}{\Psi_T}=-
 \sum_i \frac{\hbar^2}{2m_e} \left[  \frac{\nabla^2_i D}{D} -\nabla_i^2 U+ \left( \nabla_i U \right)^2 -2\frac{\nabla_i D}{D} \nabla_i U \right] +
V
\eeq
where $V$ is the potential energy of the $N$-electron system and $m_e$ is the mass of the electrons.
The total energy per particle  for this trial state, $E^{VMC}_N$, is then given by
$E_N^{VMC}=\langle E_L(\Rvec) \rangle_T/N$
where $\langle \dots \rangle_T = \int d\Rvec  \dots \Psi_T^2/ \int d\Rvec  \Psi_T^2$, and $\Rvec=(\rvec_1,\dots,\rvec_N)$
are the electronic coordinates. Performing a partial integration
 we get
\bea
E_N^{VMC}&=&\frac{1}{N}
\left\langle - \sum_i \frac{\hbar^2}{2m_e}
\left[  \frac{\nabla^2_i D}{D} -( \nabla_i U)^2  \right] + V \right\rangle_T.
\elabel{ENSJW}
\eea

The trial energy provides an upper bound to the true ground state energy which can be improved by projector Monte Carlo methods. Due to the sign-problem, it is in general impossible to project out precisely the exact Fermion ground state but within the fixed-node approximation, we can find the best energy within the nodes of the trial function. 
The fixed-node ground state can be written
\beq
\Psi_{FN}=De^{-U_{FN}}
\elabel{PsiFN}
\eeq
where $U_{FN}$ is an optimal symmetric many-body function. 
Whereas the pure distribution can be obtained by Reptation Monte Carlo methods \cite{saverio}, 
diffusion Monte Carlo  algorithms (DMC) sample only 
 the mixed distribution $\Psi_{FN}\Psi_T$,
$\langle \dots \rangle_{DMC}=\int d\Rvec  \dots \Psi_T \Psi_{FN}/\int d\Rvec \Psi_T\Psi_{FN}$,
and the (unbiased) estimator of the fixed-node DMC energy corresponding to \eq{ENSJW} can be shown to be given by:
\beq
E_N^{FN}= \frac1N \langle E_L(\Rvec)\rangle_{DMC}
=\frac1N
\left\langle -\sum_i \frac{\hbar^2}{2m_e} 
\left[  \frac{\nabla^2_i D}{D} - ( \nabla_i U)^2 - \nabla_i U \nabla_i [ U_{FN}-U] \right] +
V \right\rangle_{DMC}.
\elabel{ENDMC}
\eeq

In the following we will analyze separately the different terms in \eq{ENSJW} and \eq{ENDMC} according to their
underlying single or two-particle character. Shell effects due to the occupation of the orbitals in the Slater determinant
are mainly contained in the term involving $\nabla^2 D$, whereas the other terms of the kinetic energy have a two-body
character similar to the potential energy.

\subsection{Single particle corrections ---Shell effects}

Let us assume that the orbitals in the Slater determinant are determined  from a single particle Schr\"odinger equation,
e.g. from Kohn-Sham DFT calculations. Writing the determinant of the many-body wavefunction
\bea
D&=& \det_{in} \varphi_{ni}
\elabel{SlaterDet}
\eea 
where $\varphi_{ni}\equiv \varphi_n(\qvec_i)$ is the square matrix formed from orbital functions $\varphi_n$ evaluated at generalized electron coordinates  $\qvec_i$. For wavefunctions which do not include backflow, $\qvec_i=\rvec_i$.
We assume that the $N$ orbitals $\varphi_n$ are solutions of a Schr{\"o}dinger equation 
\beq
\left[ -\frac{\hbar^2 \nabla^2}{2m_e} +v_{\text{eff}}(\rvec) \right]\varphi_n(\rvec)=\varepsilon_n \varphi_n(\rvec)
\label{veff}
\eeq
for some effective potential $v_{\text{eff}}(\rvec)$. It follows that 
\bea
-\frac{\hbar^2}{2m_e D }
\sum_i \nabla_i^2 D 
&=& \left[ \sum_n  \varepsilon_n- \sum_i  v_{\text{eff}}(\rvec_i)  \right] .
\elabel{kineticE}
\eea
In the thermodynamic limit, the discrete summation over
energy levels will be replaced  by an integral over the density of states
\beq
\frac1N \sum_n \varepsilon_n \rightarrow \frac{1}{\rho} \int_0^{\varepsilon_F} d\epsilon \, \epsilon ~\nu_{\text{eff}}(\epsilon) 
\eeq
Here, $\nu_{\text{eff}}(\epsilon)=(2\pi)^{-d}\sum_n \int d\kvec \delta(\epsilon-\varepsilon_{n\kvec})$ is the density of states of the effective Schr\"odinger equation,
$\varepsilon_F$ is the single particle Fermi energy, and $\rho=N/\Vol$.

For metallic systems, when $\nu_{\text{eff}}(\epsilon)$
is non-vanishing for $\epsilon\approx \varepsilon_F$,  the sharp edge of the integration at the Fermi level will give rise to  large size effects, the so-called shell effects.
They can be reduced by using twist averged boundary conditions,
$\Psi(\dots,\rvec_i+L,\dots)= e^{i \vartheta L} \Psi(\dots,\rvec_i,\dots)$
where $\vartheta$ is a phase vector with $-\pi/L < \vartheta \le \pi/L$ in each direction \cite{Lin01}. 
The single particle energies in \eq{kineticE}  then depend on $\vartheta$ and only  
$N$ orbitals with lowest energies are occupied, e.g. only plane waves of smallest
wave vectors $\kvec_i+\vartheta$ are occupied for an isotropic Fermi gas.
By averaging the final
values over all twist angles
the sum becomes approximately  equal to $\int d \varepsilon$
obtained in the thermodynamic limit.
More generally, imposing the twist in the single-particle Schr\"odinger equation shifts the energies of the orbitals and twist averaging reduces the oscillatory behavior in the kinetic energy by more than an order of magnitude. Furthermore, it  restores isotropy
in the mean value of other observables such as the electron-electron interaction, leading to a more regular behavior in those quantities. 
However, in the case of a many-body calculation, twist averaged boundary conditions (TABC) do not automatically introduce a sharp Fermi surface for metallic states since only the $N$ orbitals with lowest single particle energies for any twist  are used,
 so that the thermodynamic limit of non-interacting electrons is not exactly reproduced. 
 
In order to obtain the exact single particle energy we can use grand-canonical TABC (GC-TABC) \cite{fse,momk2d}.
There, we occupy only orbitals below the Fermi energy $\varepsilon_F$  in the Slater determinant,
 $\varepsilon_n(\vartheta) \le \varepsilon_F$ but the number of electrons will then depend on the
twist angle.  GC-TABC not only reproduces exactly the  non-interacting
kinetic energy, but also the non-interacting static structure factor at all reciprocal lattice vectors commensurate with the simulation box.
Knowledge  of the single particle Fermi energy is
needed for GC-TABC to determine the mean density.  
The occupation of orbitals can be obtained by imposing
 the Fermi surface  on the single particle energies of  
 the effective Schr\"odinger equation. This Fermi energy (and therefore the mean density)  is obtained purely by the single particle effective Schr\"odinger equation converged
in the number of twist angles (or k-points).

To implement grand-canonical twist averaging for QMC calculations of charged fermions, 
we have to add an additional homogeneous background charge to ensure charge neutrality of the total system for
any given twist angle $\vartheta$.  
Of course, after twist averaging the system would be neutral, but adding a neutralizing charges is a bookkeeping exercise needed if only a finite number of selected twist angles is used. 

In order to reproduce the sharp Fermi surface within GC-TABC, a very fine mesh of twist angles has to be used.
In a system with translation symmetry such as the uniform electron gas, finite mesh errors can be completely avoided by noting that for any finite number of particles,
changes of the twist angle within a finite region, a so-called  ``pocket'',  only introduces a phase shift in all orbitals corresponding to
a change of the total momentum, 
$\Psi_{\vartheta+\delta \vartheta}(\Rvec) = \Psi_{\vartheta}(\Rvec)e^{i \delta \vartheta \cdot \sum_j \rvec_j}$.
Since the sampling weight $\propto |\Psi_{\vartheta}(\Rvec)|^2$ is unaffected by this change,
any property inside one pocket  can be calculated from the calculation of a single twist angle in the pocket 
with a weight proportional to the volume of the pocket \cite{fse}. 
These weights can be computed prior to the actual many-body simulation.
We can also use this technique for periodic solids. As in the
fully translational invariant system, the pockets are defined by the regions where 
the phase of the wavefunction changes continuously.
However, computing the different pockets introduces some overhead in the calculation, and in the following
we will discuss a simpler but equally effective reweighting method to reduce the error of using a finite mesh of twist angles.

For TABC calculations with fixed number of particles and given mesh size, performing calculations with neighbouring twist angles via reweighting  amounts in leading order to correcting the single particle
kinetic energies. Therefore, the difference between a TABC calculation done  with fixed number of twist-angles, $N_\theta$, 
and the integration over all twists will be dominated by the single particle expression
\beq
\left[ \int d^d \vartheta - \frac{1}{N_\vartheta} \sum_\vartheta \right]
 \sum_{n=1}^N  \varepsilon_n(\vartheta)
 \eeq 
As long as one uses a fixed particle number for all twists,
these corrections remain smooth. 
Similarly, we can correct for the mesh error of GC-TABC calculations by imposing the single particle Fermi surface.
In practice, the sharp Fermi surface dominates the size effects, so that we should correct the TABC results
by imposing a sharp Fermi surface giving a single particle energy correction of
\beq
\Delta T_{TABC}=
\frac{1}{\rho} \int_0^{\varepsilon_F} d\epsilon \, \epsilon~ \nu_{\text{eff}}(\epsilon) 
- \frac{1}{N_\vartheta N} \sum_\vartheta 
 \sum_n \varepsilon_n(\vartheta)
  \elabel{dTABC}
\eeq
where  the summation on the r.h.s. goes over all wave vectors of the plane wave orbitals in the TABC or GC-TABC determinant.
Adding $\Delta T_{TABC}$ to energies obtained from TABC calculations at fixed $N$, one rapidly approaches the mesh-corrected
GC-TABC results as shown in the examples below.  

For electrons in an external periodic potential created by the crystal ions, one should expect that the effective potential will
have the same periodicity as the lattice. However, in the case of a disordered potential, e.g. a two component liquid \cite{ceimc},
the potential will not be periodic. At any given system size, $N$, periodic or twisted boundary conditions still impose 
a periodicity due to the finite size of the simulation box. 
To estimate thermodynamic limit corrections the use of TABC or GC-TABC is nevertheless useful and often essential. 

Indeed for electrons in a disordered medium \cite{Mazzola,Carlotofinish}, the Fermi surface is, in general, destroyed by the external potential,
such that the lifetime of quasi-particle states  remains finite even at the Fermi surface. As a result, the sharp discontinuity 
of the momentum distribution at the Fermi surface gets smeared out. Although remaining continuous, the momentum distribution 
may still have a pronounced change in the slope very close to the Fermi surface, in particular if the disorder only weakly affects the electronic properties. The resulting oscillations in real space will eventually decay exponentially at a length scale of the mean free path of the electrons. For a disordered metallic system, this length scale can exceed the correlation length of the external disorder potential by orders of magnitudes. Nevertheless, the calculation of the mean-field path does not necessarily require large system sizes, apart from situations where one may be close to a continuous, localization driven metal-insulator transition. This surprising result can be understood considering simple potential scattering: scattering phase shifts can be reliably calculated in finite systems that are much smaller than what is 
needed to resolve a very sharp slope in the momentum distribution due to Fermi statistics. 
Stated differently, twist averaging can greatly reduce size effects even in disordered systems as long as the modification of the density of states due to disorder remains sufficiently smooth.

\subsection{Corrections arising from two-particle correlations}

\subsubsection{Potential energy correction}

Let us start considering the interaction energy in \eq{ENSJW}, again, for simplicity, written down for a one component system 
\beq
\left\langle V_N \right\rangle
=
\frac{1}{2 \Vol} \sum_{\kvec \ne 0 }
v_k 
 \left[ S_N(k)-1\right]
\eeq
The finite size error of the potential energy is
\bea
 \left[ \lim_{N \to \infty} \langle V_N \rangle  \right] -  \langle V_N \rangle 
&=& 
\frac{1}{2} \int \frac{d\kvec}{(2\pi)^d}
v_k 
 \left[ S_\infty(k)- 1 \right]
-
\frac{1}{2 \Vol} \sum_{\kvec \ne 0 }
v_k 
 \left[ S_N(k)-1\right]
\eea
Note that we have assumed that $v^N_k=v^{\infty}_k$; we use Ewald image method for the  potentials for  periodic boundary conditions \cite{Ewald}.  
If the integrand were an analytic function for all $k$, the finite size error would vanish exponentially with system size (see appendix A).
In the rest of this section we explicitly consider the case of the 3D Coulomb potential but the method can be extended to different interactions and systems of reduced spatial extensions. 

Assuming $S(k \to 0)=0$,  the leading order size correction is given by the  Madelung
constant, $\Delta v_M$, 
\beq
\Delta v_M = -\left[ \int \frac{d^3\kvec}{(2\pi)^3} -  \frac{1}{\Vol} \sum_{\kvec \ne 0} \right] \frac{v_k}{2} 
\eeq
For the 3D Coulomb potential, we have $\Delta v_M \sim N^{-1/3}$ where the proportionality factor  depends only on the geometry of the simulation box.
For a multi-component,  charge-neutral system, this term vanishes, but it must be considered in the case of GC-TABC where
a homogeneous background charge may be needed to assure charge neutrality. The remaining term of the potential energy corrections is
\bea
\Delta V_N&=& \frac{1}{2} \int \frac{d\kvec}{(2\pi)^d}
v_k  S_\infty(k)
-
\frac{1}{2 \Vol} \sum_{\kvec \ne 0 }
v_k
 S_N(k)
 \elabel{dVNdef}
\eea

Non-analytical points of the integrand will  give rise to slowest convergence of the integration.
Potential non-analytical behavior is around singularities of the potential, edges of the integration region,
$k \to 0$ and $k \to \infty$ and values $k = n k_F$ with integer $n$. 
From the local energy, we can see
that the singular behavior at $k=0$ also determines the limiting behavior of the Jastrow potential and the structure factor,
in particular, $S(k) \sim u_k^{-1} \sim v_k^{-1}  \sim k^{2}$ in $d=3$ dimensions \cite{Gaskell,ceperley78}. 
The next-to-leading order corrections beyond the Madelung corrections  are then related to the long wavelength plasmon excitation 
$\Delta V_N = \Delta V_{LO} + o(N^{-1})$ with \cite{fse}, 
\beq
 \Delta V_{LO} \equiv
\frac1N
 \frac{\hbar \omega_p}{4}
\eeq
where $\omega_p=(\rho v_k k^2/m)^{1/2}$ is the plasma frequency \footnote{The plasma frequency is independent of the wave vector in 3D, whereas it is $k$-dependent in 2D.}. 
Half of the plasmon zero point energy, $\hbar \omega_p/2$,
is actually potential energy, the missing other half is recovered from the kinetic energy \footnote{For dilute Bose gases and liquid helium we expect $u_k \sim k^{-1}$ and $S(k)\sim k$. The corresponding two-body corrections of the potential and kinetic energy are of order $N^{-4/3}$.}.
Subleading corrections  \cite{fse,drummond,momk2d,momk3d} may also be deduced by 
integrating asymptotic expansions of the structure factor around $k=0$, taking
only into account the contributions from the volume element around the origin 
in \eq{dVNdef}.
In the following, we will go beyond such an asymptotic analysis,
proposing a general and practical method to evaluate \eq{dVNdef} 
for the thermodynamic limit estimation using only results for a calculation at a finite size. 

Our best {\it a priori } choice for $S_\infty(k)$ consists of interpolating the values of $S_N(k)$ from the discrete grid in $k$-space
 to all $k$-values. From this interpolated function we can calculate the
difference between summation and integration. However, since $v_k$ is a slowly decaying function, this is not straightforward.
Since the noise of the structure factor is amplified by the volume element at large wave vector, one has to confine the integration to medium or small wave vectors. Technically, this can be achieved by splitting the potential into short and long-range parts: $v_k=v_k^{sr}+v_k^{lr}$.
Assuming an isotropic short range potential with $v_{sr}(r\ge r_c)  =  0$ for some cut-off radius $r_c$,  the
long-range contribution is then given by $v_k^{lr}=v_k-v_k^{sr}$, where $v_k^{sr}=\int d\rvec e^{-i \kvec \cdot \rvec} v_{sr}(r)$. This splitting can be done for arbitrary potentials
in an optimal way \cite{opt3d,opt2d},
such that $v_k^{lr}$ is  a rapidly vanishing  function for increasing $k$. In the case of Coulomb interaction, the short and long-range part can
also be separated  using the method introduced by Ewald \cite{Ewald,Tildesley}. Note that in the following section we will also use this procedure on the Jastrow factor, also a 
long-ranged function. The optimal split-up is routinely used in the QMC algorithms in order to  compute rapidly the potential and kinetic energy of long-ranged interactions and wavefunctions during the Monte Carlo random walk.

Let us write the potential energy per particle in terms of this breakup
\beq
\langle V_N \rangle =\frac\rho2 \int_0^{r_c} d\rvec~ v_{sr}(r)~ \left[ g_N(r)-1 \right] + \frac{1}{2\Omega} \sum_{\kvec \ne 0} v^{lr}_k [S_N(k)-1]
\eeq
where we have introduced the pair correlation function of the $N$-particle system  
\beq
g_N(\rvec)= 1+ \frac{1}{N} \sum_{\kvec \ne 0}e^{-i \kvec \cdot \rvec} \left[  S_N(\kvec)-1 \right] 
\eeq
The natural prolongation of the pair correlations to the thermodynamic limit is by assuming $g_\infty(r) \simeq g_N(r)$ for $r\le r_c$
and interpolating $S_N(k)$ to a dense grid in $k$-space, $S_\infty(k) \simeq \widetilde{S}_N(k)$.
The remaining size corrections for the potential energy are then exclusively expressed in terms of long-range contributions
\beq
\Delta V_{lr} = \left[ \int \frac{d\kvec}{(2\pi)^d} -\frac{1}{\Vol} \sum_{\kvec \ne 0} \right] \frac{v_k^{lr}}{2} \widetilde{S}_N(k)
\elabel{dVN}
\eeq 
By construction, the integration and summation of the r.h.s .of \eq{dVN} do not depend on the upper integration/ summation limit as $v_k^{lr}$
is zero for large $k$ by construction.  
In practice, we use $r_c=L/2$ together with a cubic spline interpolation of $S_N(k)$
to continue the values on the discrete $k$-grid to the continuum
 to obtain $ \widetilde{S}_N(k)$. 
 For a multicomponent system interacting only via Coulomb forces, only the charged structure factor is needed for the potential energy
 and we impose the boundary conditions: $S(0)=\left(dS/dk\right)_0=0$.
 
 Notice, that for the derivation of \eq{dVN} we have assumed that the short range part of the pair correlation function remains
 unchanged in the thermodynamic limit. This would be the case if the structure factor was an analytical function of $k$ which is in general not the case. Although $S(k) \sim k^2$ for 3D, non-analytic behavior $\sim k^3$ is expected beyond leading order  giving rise to additional corrections of order $N^{-2}$ which we neglect in the following
 \footnote{Here, the discussion is limited to 3D electronic systems. In 2D or 1D,
the non-analytical behavior of the structure factor are more important and $\Delta V_{sr}$ 
together with the corresponding corrections for the kinetic energy $\Delta T_{U}^{sr}$ should be added for size extrapolation. See appendix D.}.
In appendix D we describe how to go beyond this assumption to include sub-leading corrections
due to non-analyticities in the structure factor involving also the short-range part of the potential and   
explicitly show how to perform the calculations with the Ewald potential for cases where the optimized potentials are not available.

\subsubsection{Kinetic energy correction}
Let us now consider the kinetic energy contribution involving the Jastrow correlations, the remaining term of \eq{ENSJW},
\beq
T_U= \frac1N \left\langle  \sum_i \frac{\hbar^2}{2m_i} [\nabla_i U]^2 \right\rangle
\eeq
Again, we analyze the expression in terms of the Fourier components. Restricting to a single component system
with $U= \frac1{2\Vol} \sum_\kvec u_{\kvec} \rho_{\kvec}\rho_{-\kvec}$, we have
\bea
T_U &= &- \frac{\hbar^2}{2m_eN}  \frac{1}{\Vol^2}  \sum_{\kvec \ne 0, \kvec' \ne 0} (\kvec \cdot \kvec') u_\kvec u_{\kvec'}
\rho_{\kvec+\kvec'} \rho_{-\kvec} \rho_{-\kvec'}\\
&\simeq& 
 \frac{1}{\Vol} \sum_{\kvec\ne 0}  \frac{\hbar^2 k^2}{2 m_e} \rho u_{\kvec}u_{-\kvec} S_N(\kvec)   
 \elabel{TU}
\eea
where we have neglected all terms with $\kvec \ne -\kvec'$  corresponding to the RPA approximation which
becomes exact in the long wavelength limit \cite{PinesNozieres}.

In order to analyze the finite size corrections for the energy
we interpolate the Jastrow potential, $u(\kvec)$, from its values at discrete $\kvec$-points to all values of $k$. As with the potential energy, the kinetic energy error from pair correlations,  
$\Delta T_U =  \left[ \lim_{N \to \infty} \langle T_U \rangle /N \right] -  \langle T_U \rangle /N$, reduces to 
an integration error.
The non-analytical behavior of the integrand around $k=0$ for Coulomb systems
gives rise to slow convergence of the integration; the leading order is  given by the plasmon contribution \cite{fse}.
\beq
\Delta T_U^{LO}=\Delta V_{LO}
=\frac1N
 \frac{\hbar \omega_p}{4}
\eeq
To go beyond leading order, we split the long-range from the short range part of the Jastrow potential, 
$u_k=u_{sr}(k)+u_{lr}(k)$, so that 
we arrive at the following expression
\beq
\Delta T_U^{lr}=
\left[ \int \frac{d\kvec}{(2\pi)^d} -  \frac{1}{\Vol} \sum_{\kvec\ne 0} 
\right]
 \frac{\hbar^2 k^2}{2 m_e} \rho u_{lr}(k)\left[ 2 u_k-u_{lr}(k) \right] \widetilde{S}_N(k)   
 \elabel{dTU}
 \eeq
As the integrand on the  r.h.s. vanishes rapidly with $k$, we only have to interpolate  the structure factor at small $k$
and work out the corrections similar to those for the potential energy.

\subsubsection{Backflow corrections} 

The above corrections are for a Slater-Jastrow wavefunction.
Backflow wavefunctions considerably improve the accuracy of QMC calculations \cite{Lee,Kwon3D,bf,BFNeeds,BFh,manybody}
and have been generalized to systematically approach the ground state energies \cite{iteratedBF}.
Let us consider that the orbitals $\varphi_{ni}=\varphi_n(\qvec_i)$ inside the Slater determinant, \eq{SlaterDet}, 
 are built using general backflow coordinates, $\qvec_i=\rvec_i + \etavec_i$ where $\etavec_i$ is a function of all other coordinates.
The derivatives of $\etavec_i$ will then give rise to additional terms of the kinetic energy with corresponding finite size corrections
which have not discussed so far. Using RPA like arguments, we can estimate the dominating terms in the laplacian
of the kinetic energy (see appendix C)
\beq
\left\langle -\frac{\hbar^2}{2m_e} \frac{\nabla^2 D}{D} \right\rangle \approx \left[1+\frac{1}{\Vol}\sum_\kvec s_N(k)\right] \left\langle -\frac{\nabla_q^2D}{2m_eD} \right\rangle
= \left[1+\frac{1}{\Vol}\sum_\kvec s_N(k)\right] \left[ \sum_n \varepsilon_n - \left\langle \sum_i v_{\text{eff}}(\qvec_i)\right\rangle \right]
\eeq
with
\beq
 s_N(k) = \frac{k^2y_k}{d} \left[ 2+ \rho k^2 y_k - \left( 2-\rho k^2 y_k \right) S_N(k) \right]
 \elabel{sNk}
\eeq
where $d$ is the spatial dimension and the backflow potential, $y_q$,
is related to the quasiparticle coordinates \cite{BFminus} via $\etavec_i= \frac{i}{\Vol} \sum_q \qvec y_q (e^{i \qvec \cdot \rvec_i} \rho_{-\qvec}-1)$.

As before, we can now derive the size corrections of the kinetic energy due to backflow, $\Delta T_{BF}$, using
\bea
\Delta T_{BF} &\simeq& t 
\left[ \int \frac{d\kvec}{(2\pi)^d} -  \frac{1}{\Vol} \sum_{\kvec\ne 0} 
\right] \widetilde{s}_N(k)  \elabel{dTbf}\\
t &=& \frac{1}{N} \left[  \sum_n \varepsilon_n - \left\langle \sum_i v_{\text{eff}}(\qvec_i)\right\rangle  \right]
\label{TBF}
\eea
where $\widetilde{s}_N(k)$ is given by \eq{sNk} where the long-range part of  $y_k$ and $y_k^2$ together with
an interpolation of the structure factor is used as was done with the previous kinetic energy corrections without backflow.

From the long-range limit of the electron electron backflow \cite{bf}, 
we can estimate the leading order size effects of backflow for a metallic system in 3D
\beq
\Delta T_{BF}^{LO}= - \frac{t}{3 N} 
\label{TBFLO}
 \eeq
 where $t$ is the single particle kinetic energy, $t \simeq 3 k_F^2/10m_e$ for a system with an isotropic Fermi surface.

\subsection{Projection Monte Carlo methods and mixed estimators}

Starting from a trial wavefunction, the true ground state wavefunction 
can be sampled using Projector Monte Carlo methods. Using 
the fixed-node approximation for fermions to circumvent the sign problem,
the optimal ground state wavefunction constrained by the nodes of the
given antisymmetric Slater determinant, $D$, is determined, \eq{PsiFN}.
When using twisted boundary conditions for TABC and GC-TABC, we replace the fixed-node procedure with the
fixed phase approximation \cite{FixedPhase}.
Extrapolating the calculations imposing the nodes/phases obtained from the same effective potential
will therefore lead to identical single-particle size corrections as within VMC,  \eq{dTABC},
as these shell corrections are due to the behavior of the phase of the many-body wavefunction
which is unaffected by the restricted random walk of the projection.

Concerning the two-particle corrections, size-corrections beyond shell effects are directly related to the bosonic long wavelength modes,
so that projection may lead to essential changes. Within projection Monte Carlo methods \cite{RMP,saverio}, the total energy
of the system is most easily obtained  by commuting the Hamiltonian to one end of the path where the energy can be
obtained from the mixed estimator, \eq{ENDMC}. In the case where the exact long-range behavior of
the Jastrow function is already imposed in the trial wavefunction, we can neglect size effects
of the term involving $U_{FN}-U$, and two-particle size effects for the total energy are corrected
by adding potential and kinetic energy corrections, \eq{dVN} and \eq{dTU}, where $\widetilde{S}_N$ is obtained
from the mixed estimator of the structure factor. Whereas the total energy estimator is unbiased, separating
potential from kinetic energy contributions may be biased in this procedure.

Unbiased calculation of the structure factor can be done using  reptation Monte Carlo \cite{saverio}, 
so that we can directly apply the VMC formulas for potential and kinetic
energy corrections, \eq{dVN} and \eq{dTU}, respectively, 
as long as the exact long-range behavior of the Jastrow is contained in the trial wavefunction.
Although reptation Monte Carlo can be extended to obtain unbiased estimators for off-diagonal quantities such as the 
momentum distribution \cite{momk3d}, it is simpler to determine the kinetic energy as the
difference of the total energy and the potential energy.

In cases where the trial wavefunction does not contain the correct long-range behavior, the term involving
 $U_{FN}-U$ becomes relevant for the size extrapolation. To estimate this correction, we can use a quite general
 relation between the structure factor and the effective Jastrow \cite{Gaskell}, $u_{\text{eff}}(k)$, valid in the long wavelength limit
 \beq
 S^{-1}(k)=S_0^{-1}(k)+2\rho u_{\text{eff}}(k)
 \elabel{skform}
 \eeq
 where $S_0(k)$ is the ideal gas structure factor (see appendix B). For metallic systems,  its contribution to the effective Jastrow factor is negligible, e.g.
 for the DMC mixed estimator of the structure factor, we have
 \beq
 S_{DMC}(k)=\frac{1}{\rho [u_{FN}(k)+u(k)]}, \quad k \to 0
 \elabel{SDMC}
 \eeq
 and we can obtain $u_{FN}(k)$ from the mixed DMC estimator, as  the Jastrow factor of the trial wavefunction, $u(k)$, is known. Knowledge of $u_{FN}$, allow us to calculate the two-particle 
 size corrections of the total energy from \eq{dVN} and \eq{dTU} together with a similar term to take into account the 
 corrections involving
$U_{FN}-U$  in \eq{ENDMC}.

Let us discuss explicitly the common practice of using a short range function for the electron-electron Jastrow factor.
In this case the long-range part of $\lim_{k\rightarrow 0} u(k)\rightarrow const$ compared to the exact one which diverges as $k^{-2}$.
Whereas the kinetic energy contribution, $\eq{dTU}$, also remains negligible, the potential energy contribution
with the use of the mixed estimator $S_{DMC} \simeq [\rho u_{FN}(k)]^{-1}$ is twice as large as that obtained by 
using the exact structure factor $S(k) \simeq [2\rho u_{FN}(k)]^{-1}$. 
We see, that the leading order corrections for the total energy
are indeed correctly obtained by the mixed estimator, 
however, the true potential energy as obtained by the use of the unbiased estimator  contributes only half of this correction. The other half comes from the kinetic energy.
Notice that the exact structure factor also
agrees with the one obtained from the extrapolation
formula $S(k) \simeq S_{EXTR}(k) \equiv 2 S_{DMC}(k) - S_{VMC}(k)$ as long as $u_{FN}(k)-u(k)$ is small. 
In practice, for calculations based on a short range electron-electron Jastrow,
we can use the interpolation of $S_{DMC}(k)$ in the potential energy extrapolation to obtain
the total energy size corrections and that of $S_{EXTR}(k)$ to obtain the potential energy corrections. From the difference
we can estimate the kinetic energy corrections without knowledge of the exact long-range Jastrow correlations.

Therefore, although it is not necessary to include the correct long-range behavior into the trial wavefunction, the long-range correlations affact the size extrapolation of DMC energies.
Correct sampling of these long wavelength correlations 
requires long projection times, since the projection time scales as $L^2=N^{2/3}$ and  long projection times require a larger population of walkers to remain unbiased.

Size extrapolations are frequently based on VMC energies, only. Without including long-range contributions
to the Jastrow, there is no guarantee that these size extrapolations can be transferred to the DMC energies. Still,
as long as the range of the short range Jastrow potential used in the trial wavefunction
grows proportional to the system size, the optimization
of a sufficiently flexible functional form will converge to the correct long-range behavior and allow a correct estimation
of the size corrections.

\subsection{Extension to inhomogeneous densities}
\label{inhomo}

The above formulas have been derived for the case of a homogeneous one-component system. Here, we briefly discuss
how they should be used for inhomogeneous electronic densities. Define $\overline{\rho}_\kvec \equiv \langle \rho_\kvec \rangle$ and separate the mean-values from the fluctuating quantities,
$\delta \rho_{\kvec} \equiv \rho_{\kvec}-\overline{\rho}_\kvec$,
in the structure factor, $S_N(\kvec) =\overline{S}_N(\kvec)+\delta S_N(\kvec)$ with
\bea
\overline{S}_N(\kvec) &=& \frac{1}{N} \overline{\rho}_\kvec \, \overline{\rho}_{-\kvec} \\
\delta S_N(\kvec) &= &\frac{1}{N} \langle \delta \rho_\kvec \delta \rho_{-\kvec} \rangle.
\eea
Extensions of $S_N(\kvec)$ to the thermodynamic limit needed for the size corrections should be done separately for $\overline{S}_N(\kvec)$ and
$\delta S_N(\kvec) $. Whereas extensions of both quantities may be needed for 
the potential energy corrections, \eq{dVN}, only $\delta S_N(\kvec)$ 
enters the kinetic energy, \eq{dTU} and \eq{dTbf}. However, $\overline{S}_N(\kvec)$ 
will have the periodicity of the unit cell for crystal structures and only $\delta S_N(\kvec)$ enters the potential energy corrections.
In the following examples we will refer to $\delta S_N(\kvec)$ as the fluctuating structure factor.

\section{Examples}

In the following, we provide examples of the  finite size errors using the
correction schemes discussed above.
In general, we have performed calculations averaging over twisted boundary conditions, where the
twist angles, $\theta_\alpha \in ]-\pi/L,\pi/L]$ ($\alpha=x,y,z$), are chosen on a linear grid with $M^d$ discretization points,
$\theta_\alpha= (m/M -1/2) \pi/L$, $m=1,2,\dots M$. Bare results for the total energy per electron using TABC or GC-TABC
are denoted by $E_{TABC}$ and $E_{GC-TABC}$, respectively. We then give the remaining  shell corrections,
$\Delta T_{TABC}$, or $\Delta T_{GC-TABC}$, obtained from \eq{dTABC}.

The leading order two-particle energy correction for static ions, 
assuming the correct long-range behavior of the electronic Jastrow function, is given by the plasmon zero-point energy, $\Delta E_{LO}=\hbar \omega_p/2N=\sqrt{3}r_s^{-3/2}/(2N) \text{Ha}$, independent of the shape of the 
supercell used. Here and in the following, $r_s=(4\pi \rho_e a_B^3/3)^{-1/3}$ is the Wigner-Seitz density parameter, $\rho_e$ is the electronic density and $a_B=\hbar^2/m_e e^2$ is the Bohr radius.  Corrections beyond leading order $\Delta E_{lr} \equiv \Delta T_U^{lr}+ \Delta V_{lr}$ are calculated
by using the optimized long-range part of the corresponding potentials in \eq{dVN} and \eq{dTU}, and interpolating
$S_N(k)$ using  cubic splines and assuming that $S_N(0)=S_N'(0)=0$.

\subsection{Crystalline hydrogen: perfect crystal}

Here we consider b.c.c. hydrogen at a density of $r_s=1.31$, and we neglect the zero point motion of the protons.
We present first VMC calculations with the simplest orbitals in the Slater determinant (spherical Fermi surface),
$\phi_{n}(\rvec)=e^{i \kvec_n \cdot \rvec}$ with $k \le k_F$ where $k_F$ is the Fermi wave vector 
and fully analytical potentials including long-range parts for the many-body backflow and Jastrow potential. We separately study the
influence of backflow coordinates on the size extrapolation. 
 We then present DMC results using realistic DFT-bandstructure orbitals without backflow including only short-range Jastrow functions. The Reptation Monte Carlo method is used to obtain pure estimates of the potential energy in the latter case.

\subsubsection{Slater-Jastrow wavefunction with spherical Fermi surface}

\begin{figure}
\includegraphics[width=0.5\textwidth]{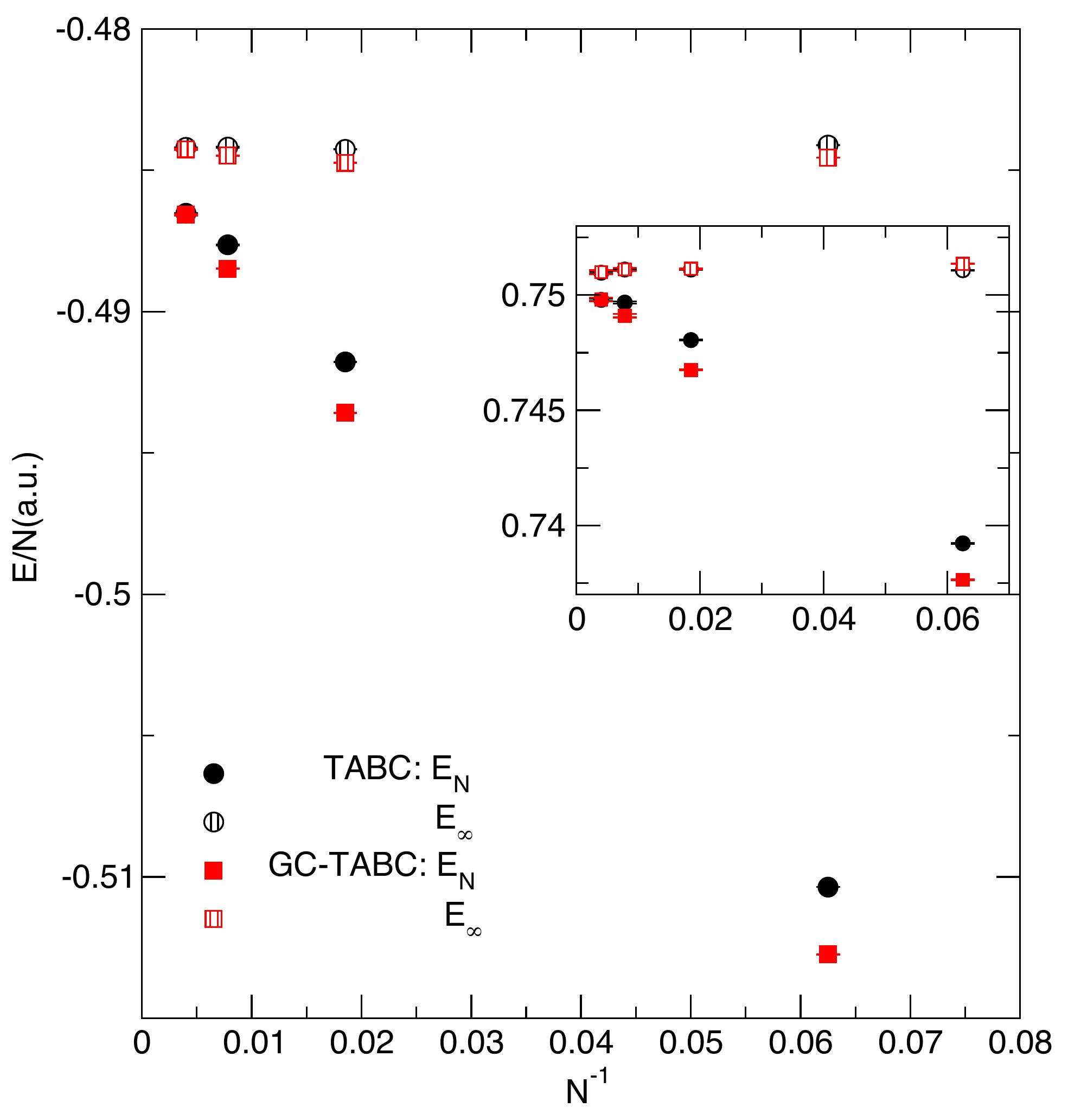}
  \caption{Energy per atom for b.c.c. hydrogen at $r_s=1.31$ using a Slater-Jastrow wavefunction with plane wave orbitals
  occupied up to the Fermi surface for various system sizes, $N$. We show the uncorrected twist averaged results, 
  $E_{TABC}$ and $E_{GC-TABC}$, together with the size corrected ones, $E_\infty(TABC)=E_{TABC}+\Delta T_{TABC}+\Delta T_{U}^{lr}+\Delta V_{lr}$ and $E_\infty(GC-TABC)=E_{GC-TABC}+\Delta T_{GC-TABC}+\Delta T_{U}^{lr}+\Delta V_{lr}$.
 In the inset we show the corresponding values of kinetic energy.
}
\label{fg_bcc_sj}
\end{figure}

The simplest trial wavefunction is a Fermi liquid state, where the Slater determinant is formed from plane waves filling the
Fermi sea up to an isotropic Fermi wave vector $k_F$. We have used an expression (analytic in k-space)
based on the RPA \cite{ceperley78,bf} for the Jastrow factor. Then the trial wavefunction for various values of N is guaranteed to approach the large N limit very smoothly. (Optimized Jastrow factors could have an additional noisy component.)
The VMC results are given in table \ref{hydrogenBCC} and Fig.~\ref{fg_bcc_sj}, we can see that convergence of the energy per particle
to less than $1$mHa  can be reached already with $N=16$ electrons. 
For  TABC, extrapolation with $M$ is smooth, but  the
use of  the shell corrections, $\Delta T_{TABC}$,  to mimic the sharp Fermi surface at $k_F$ is essential to reach
a high precision comparable to  GC-TABC results.
 
Kinetic and potential energy corrections are given separately. 
Interestingly, kinetic and potential energy corrections stemming from the two particle correlations remain equal to high precision beyond leading order, and approach the leading order expression
from below.

\begin{table}
 \begin{tabular}{|c|c||c||c|c||c|c|c||c|} \hline
 N & M & $E_{TABC}$ & $\Delta T_{TABC}$ &  $\Delta E_{LO}$& $\Delta T_U^{lr}$ & $\Delta V_{lr}$  & $\Delta E_{lr}$ & $E_\infty$(TABC)  \\\hline
16  & 161 &  -0.510358(3)  & -0.001817  &  0.036100 & 0.013667  & 0.014393  & 0.028059  &   -0.484115(3) \\
 54 &  161 & -0.491779(6) &  -0.001377  &  0.010696 & 0.004440  & 0.004455  & 0.008895  &   -0.484261(6)  \\
128 & 161 & -0.48764(2)   &  -0.000602 &   0.004512 & 0.002035  & 0.002021  & 0.004056  &   -0.48419(2)\\
250 & 161 & -0.486525(4) &  -0.000013 &   0.002310&  0.001195  & 0.001150  & 0.002345  &  -0.48417(3)\\ 
 \text{lin.extrap.}&   &    &                       &                     &                     &                     &                    & -0.4851(1)    \\
 \hline
 \hline
 N & M & $E_{GC-TABC}$ & $\Delta T_{GC-TABC}$ &  $\Delta E_{LO}$ & $\Delta T_U^{lr}$ & $\Delta V_{lr}$ &$\Delta E_{lr}$ & $E_\infty$(GC-TABC)  \\\hline
16  & 161 & -0.512739(4)   & -0.000034 & 0.036100  & 0.013744 & 0.014473 & 0.028217 &  -0.484556(4) \\
 54 &  161& -0.493581(6)    &-0.000027 & 0.010696  & 0.004424 & 0.004441 & 0.008865 &  -0.484743(6) \\
128& 161 & -0.488484(3)   & -0.000003 &  0.004512 & 0.002009 & 0.001996 & 0.004005 &  -0.484483(3)\\
250 & 81  & -0.48658(2)      &  0.000012  & 0.002310 & 0.001171&  0.001131 & 0.002302 &  -0.48426(2)\\ 
 \text{lin.extrap.} &  &      &                     &                    &                  &                     &                     &   -0.48476(2)   \\
 \hline
\end{tabular}
\caption{Energy corrections  in Hartrees per electron for b.c.c. hydrogen at $r_s=1.31$ vs. number of electrons, $N$, with various size correction estimates. Energies are computed with VMC and using a Slater-Jastrow trial function with plane waves orbitals.  
TABC: twist-averaged boundary conditions, GC-TABC: grand-canonical twist average boundary conditions.
The number of twists in each dimension was $M$,
$\Delta E_{LO}$: Leading order energy size corrections (plasmon formula), 
$\Delta E_{lr}=\Delta T_{lr}+\Delta V_{lr}$ size correction using fit of $S_k$ to compute all long-range corrections
where \eq{dTU} is used for the kinetic energy correction, $\Delta T_U^{lr}$, and \eq{dVN} for potential energy correction, $\Delta V_{lr}$. The {\it a priori} best estimate for $E_{\infty}$ using only quantities in the
$N$-particle system, denoted by $E_\infty (TABC)=E_{TABC}+\Delta T_{TABC}+\Delta E_{lr}$ for TABC and, 
similar, $E_\infty(GC-TABC)$ for GC-TABC. Linear extrapolation of $E_{TABC}$ ($E_{GC-TABC}$) using the data
for $54\le N\le 250$ is given in the lines of $N=54-250$. Statistical errors in the last digit are given in parentheses.}
\label{hydrogenBCC}
\end{table}

\subsubsection{Backflow wavefunction with spherical Fermi surface}

\begin{figure}
\includegraphics[width=0.5\textwidth]{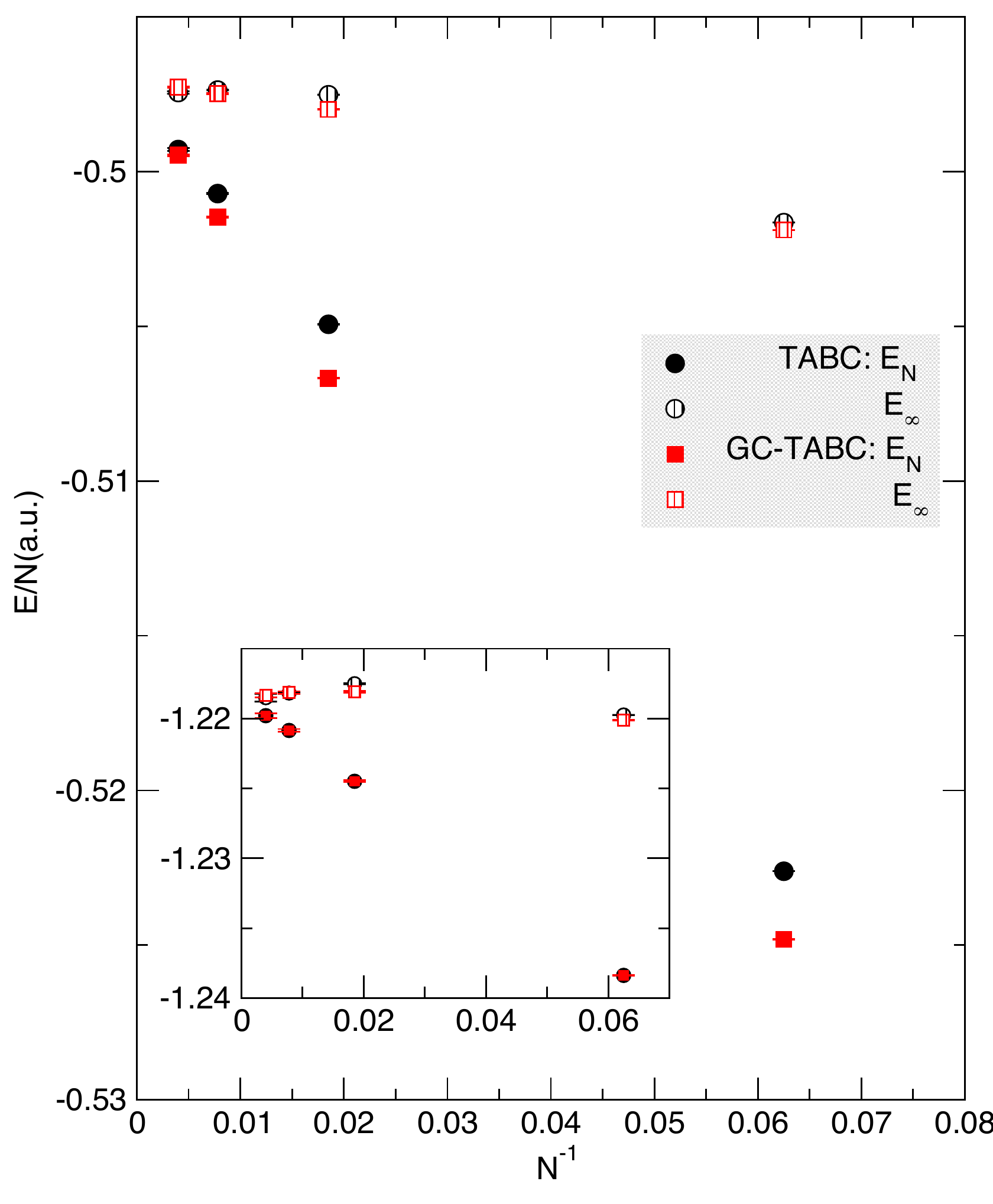}
  \caption{Energy per atom for b.c.c. hydrogen at $r_s=1.31$ using a backflow wavefunction with plane wave orbitals occupied up to the Fermi surface for various system sizes, $N$. We show the uncorrected twist averaged results, 
  $E_{TABC}$ and $E_{GC-TABC}$, together with the size corrected ones, $E_\infty(TABC)=E_{TABC}+\Delta T_{TABC}+\Delta T_{U}^{lr}+\Delta T_{bf}+\Delta V_{lr}$ and $E_\infty(GC-TABC)=E_{GC-TABC}+\Delta T_{GC-TABC}+\Delta T_{bf}+\Delta T_{U}^{lr}+\Delta V_{lr}$.
 In the inset we show the corresponding values of potential energy.
}
\label{fg_bcc_bf}
\end{figure}

\begin{figure}
\includegraphics[width=0.5\textwidth]{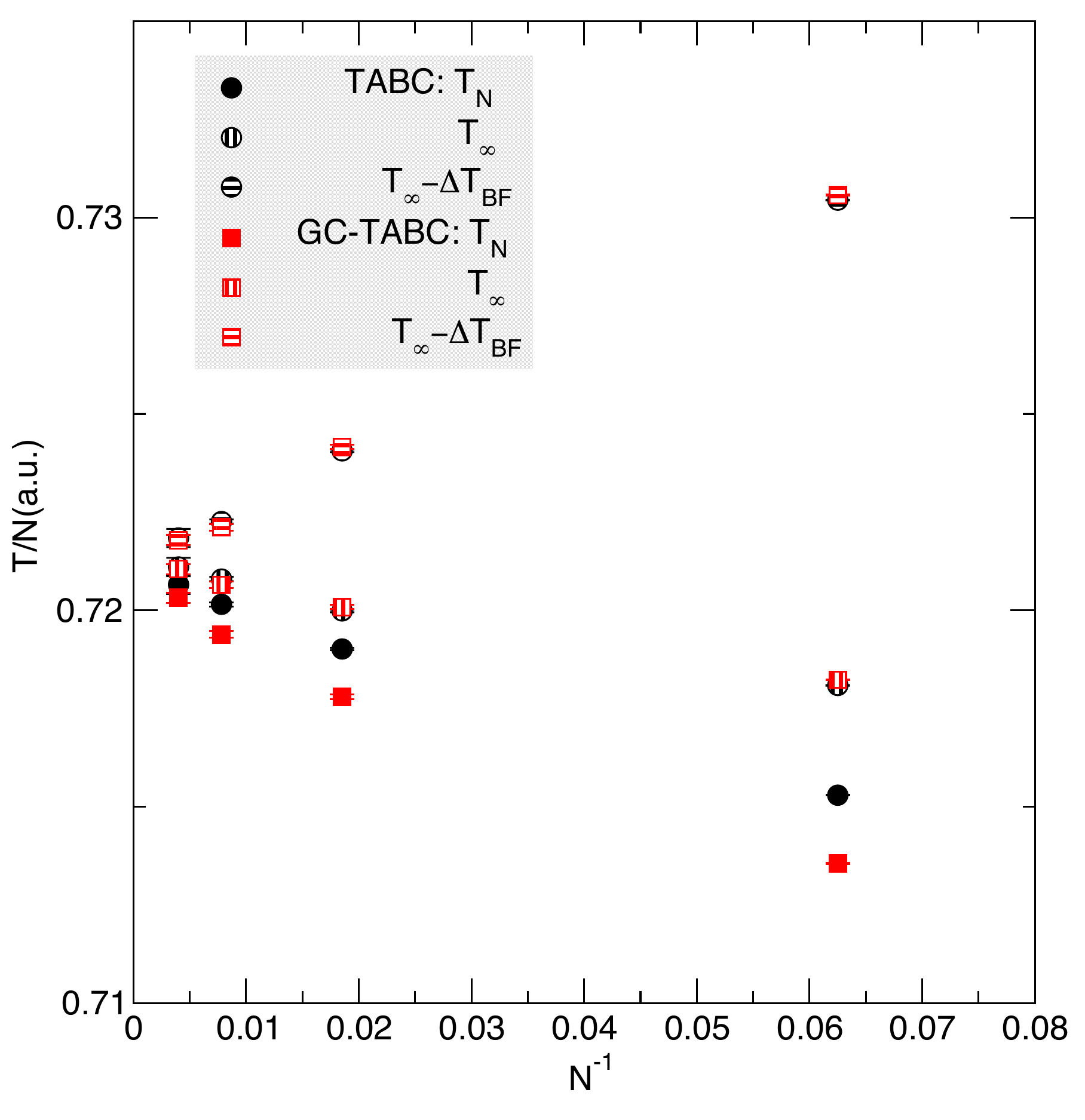}
  \caption{Kinetic energy per atom for b.c.c. hydrogen at $r_s=1.31$ using a backflow wavefunction with plane wave orbitals
  occupied up to the Fermi surface for various system sizes, $N$. We show the uncorrected twist averaged results, 
  $T_{TABC}$ and $T_{GC-TABC}$, together with the size corrected ones, $T_\infty(TABC)=T_{TABC}+\Delta T_{TABC}+\Delta T_{U}^{lr}+\Delta T_{bf}$ and $T_\infty(GC-TABC)=T_{GC-TABC}+\Delta T_{GC-TABC}+\Delta T_{bf}+\Delta T_{U}^{lr}$ with and
  without backflow corrections.
}
\label{fg_bcc_T_bf}
\end{figure}

In table \ref{hydrogenBCCBF} we show the results for the same system as before but including backflow 
coordinates to evaluate the orbitals \cite{bf}. Whereas potential energy corrections work as well as in the case without backflow (see Fig.\ref{fg_bcc_bf}),  
the explicit backflow corrections derived above, $\Delta T_{BF}$, are clearly improve the extrapolation (see
Fig.~\ref{fg_bcc_T_bf}). Still, the underlying approximations in their derivation introduce a larger error   than for the Slater-Jastrow trial function.

\begin{table}
 \begin{tabular}{|c|c||c||c|c|c||c|c|c|c||c|} \hline
 N & M & $E_{TABC}$ & $\Delta T_{TABC}$  & $\Delta T_{BF}^{LO}$ &  $\Delta E_{LO}$& $\Delta T_{BF}$ &$\Delta T_U^{lr}$ & $\Delta V_{lr}$  & $\Delta E_{lr}$ & $E_\infty$(TABC)  \\\hline
16  & 161 &  -0.522611(2)  & -0.001817 & -0.013414 & 0.036100 & -0.012354 & 0.016964 & 0.018167& 0.035132 &-0.501651(2) \\
 54 &  81 &   -0.50494(1)    &  -0.001385 & -0.003975 & 0.010696 & -0.004075 & 0.006429 & 0.006448 &0.012877 & -0.49752(1)  \\
128 & 81&    -0.50071(1)   &  -0.000604 &  -0.001677 & 0.004512 & -0.001464 &0.002712  & 0.002698 &0.005410 & -0.49737(1)\\
250 & 81 &   -0.49929(4) &    -0.000191 &  -0.000858 & 0.002310&  -0.000740 &0.001382  & 0.001395 &0.002777&  -0.49745(4)\\ 
 \text{lin.extrap.}&   &     &                       & &                    &  &                   &                     &                    & -0.49765(4)  \\
 \hline
 \hline
 N & M & $E_{GC-TABC}$ & $\Delta T_{GC-TABC}$ & $\Delta T_{BF}^{LO}$ &  $\Delta E_{LO}$ & $\Delta T_{BF}$ & $\Delta T_U^{lr}$ & $\Delta V_{lr}$ &$\Delta E_{lr}$ & $E_\infty$(GC-TABC)  \\\hline
16  & 161 & -0.524814(3)   & -0.000034 & -0.013414 &0.036100 & -0.012344 & 0.017050 & 0.018254 & 0.035304 &  -0.501888(3) \\
 54 &  161& -0.506684(7)    &-0.000027 & -0.003975 &0.010696 & -0.004078 & 0.006389 & 0.006409 & 0.012797 &  -0.497991(7) \\
128& 81   & -0.50147(2)   &   -0.000005 & -0.001677 &0.004512 & -0.001463 & 0.002732 & 0.002718 & 0.005450 &  -0.49749(2)\\
250 & 81  & -0.49948(2)      &  0.000012 &-0.000858 &0.002310&  -0.000734 & 0.001456 &  0.001131 & 0.002915 &  -0.49728(2)\\ 
 \text{lin.extrap.} &  &      &                     &  &                 & &                  &                     &                     &  -0.4976(1)    \\
 \hline
\end{tabular}
\caption{Energy corrections  in Hartrees per electron for b.c.c. hydrogen at $r_s=1.31$ for 
different number of electrons, $N$, with various size correction estimates. Energies are computed with VMC  using a Slater-Jastrow trial function where plane waves orbitals contain backflow coordinates. Additional kinetic energy corrections due to backflow are denoted by $\Delta T_{BF}^{LO}$ for the leading order formula, Eq.~(\ref{TBFLO}), and $\Delta T_{BF}$ from the interpolation of the
static structure factor together with Eq.~(\ref{TBF}). The other
symbols are defined in the caption of table \ref{hydrogenBCC}. }
\label{hydrogenBCCBF}
\end{table}

\subsubsection{Slater-Jastrow wavefunction with DFT band structure}

Next, we consider orbitals obtained from a density functional theory calculation inside the
Slater determinant using QMCPACK \cite{qmcpack1,qmcpack2}.  Single particle orbitals were obtained from Quantum Espresso using the PBE functional \cite{QE2009}.  The oribtals were generated on an 8x8x8 shifted Monkhorst pack grid using a planewave cutoff of 200Ry.  A hard Troullier-Martins pseudopotential with a cutoff of $r_c=0.5a_0$ was used to eliminate the $1/r$ divergence in the DFT calculation.

We used a Slater-Jastrow type trial wavefunctions without backflow.  The Jastrows factor consisted of a sum of radially symmetric short-ranged one and two-body terms without long-range contributions.  The cutoff radius for the short-ranged terms was chosen to be the Wigner-Seitz radius of each simulation cell.  A fully optimizable b-spline form was used for all Jastrow terms, which we optimized with variational Monte Carlo using the linear method.  

For the QMC calculations, we considered supercells with $N=16,128,1024$ atoms.  Twist averaging was used to obtain all reported quantities.  To generate the twists, we used the same Monkhorst-Pack grid as used for orbital generation. 
 Reptation Monte Carlo (RMC) was used for the $N=16,128$ supercells to compute both mixed and pure energy estimates.  For the $N=1024$ supercell, we used Diffusion Monte Carlo to calculate the total energy.
In all calculations a time step of $\tau=0.0075Ha^{-1}$ was used.  In the RMC calculations, a projection time of $\beta=4.5Ha^{-1}$ was used. 

\begin{figure}
\includegraphics[width=0.5\textwidth]{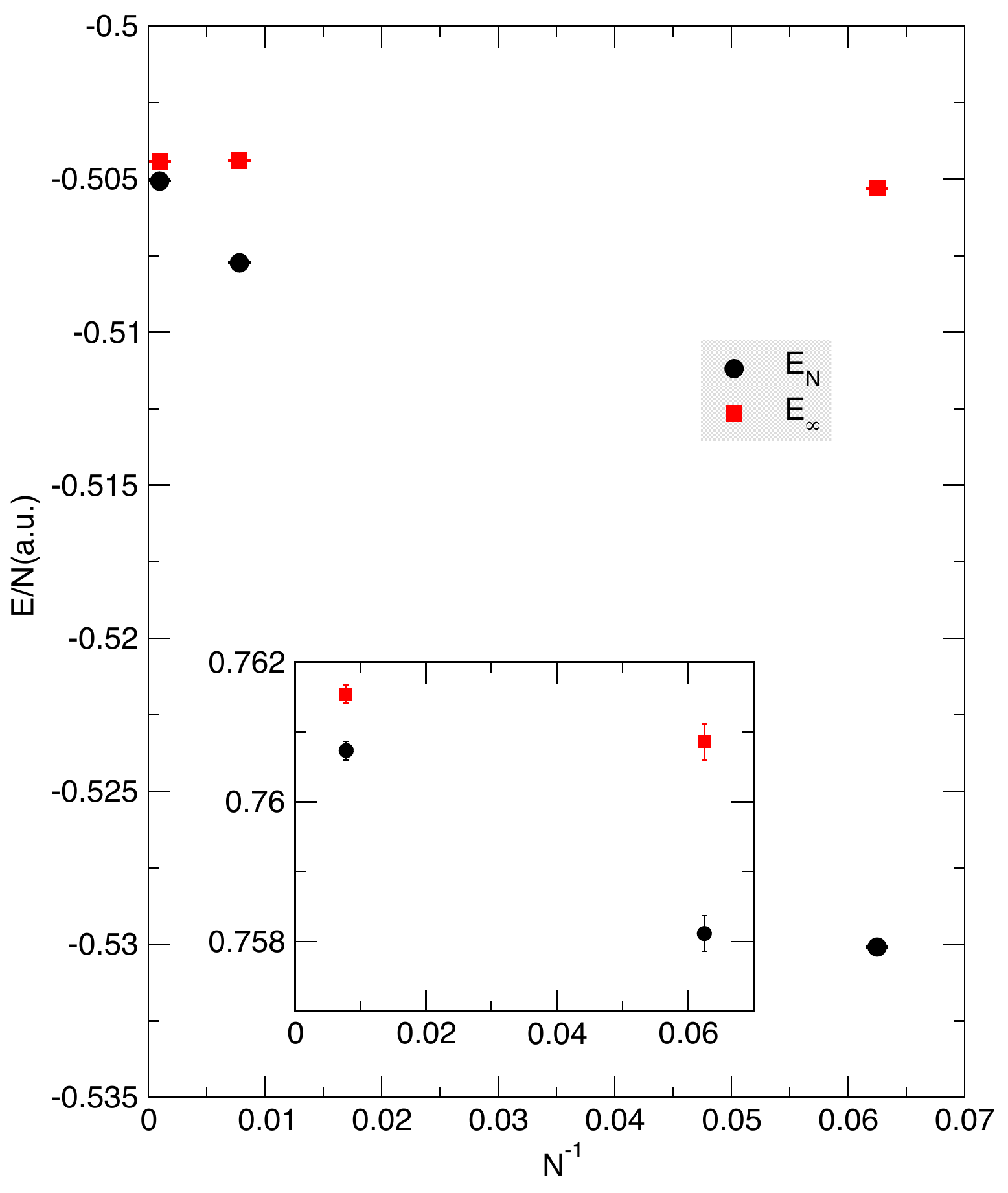}
  \caption{Energy per atom for b.c.c. hydrogen at $r_s=1.31$ using a Slater-Jastrow wavefunction with DFT orbitals
 for various system sizes, $N$. We show the uncorrected twist averaged results, 
  $E_N$, together with the size corrected ones, $E_\infty=E_N+\Delta T_{U}^{lr}+\Delta V_{lr}$.
 In the inset we show the corresponding values of kinetic energy. We also show the linear fits to the uncorrected data and 
 the corresponding extrapolated values in the thermodynamic limit.
}
\label{dft_bcc_sj}
\end{figure}

Results are summarized in table \ref{dft_bcc} and Fig. \ref{dft_bcc_sj}. 
The total energy corrections were obtained by the mixed estimator for the fluctuating structure factor whereas the potential energy corrections, $\Delta V_{lr}$ have been calculated from the pure estimator using RMC. The kinetic energy corrections, $\Delta T_{lr}$, then result from the difference of total and potential energy corrections.

We see, that the finite size error after corrections in this case of more realistic orbitals is comparable to the previous calculations using  the simple plane wave determinant, even without including long-range components in the Jastrow potential. However, the use of ``exact'' estimators was essential to reach this precision for kinetic and potential energy separately. 

\begin{table}
\centering
\label{DFTsolid}
\begin{tabular}{|c|c||c||c|c|c||c|c||c||c|c|c|}
\hline
$N$ & $M$ &$E_{TABC}$ & $T_{TABC}$  & $V_{EXTR}$ &$V$ &$\Delta T$ & $\Delta V$ &$\Delta E$  & $T_\infty$& $V_\infty$ & $E_\infty$ \\
 \hline
 16 & 8  & -0.53009(3) & 0.7581(3) & -1.2894(3)& -1.2882(3) &  0.00274 & 0.02205 & 0.02479 & 0.7608(3)& -1.2662(3) & -0.50530(3)\\
  128 & 4  & -0.50774(2) & 0.7607(1) &-1.2714(1)& -1.2685(1) &0.00081   & 0.00254 & 0.00335  &0.7615(1) & -1.2660(1)&  -0.50439(2) \\
   1024   & 2 &  -0.50507(1)  & && & &  & 0.000641  & & & -0.50443(1)\\ 
   \text{lin.extrap.} & & & &  && & &&0.7611(3) & -1.2657(3) & -0.50465(1) \\
\hline
\end{tabular}
\caption{Finite size corrections for b.c.c. atomic hydrogen ($r_s=1.31$) at zero temperature using DFT orbitals in the Slater determinant.  The potential energy per electron, $V$, was obtained using the pure estimator within RMC, whereas the kinetic energy was calculated via $T=E-V$. From the extrapolated results of the DMC calculations using the mixed estimator$V_{EXTR}=2 V_{DMC}-V_{VMC}$, we conclude that the mixed estimator introduces a bias of $3$mHa which is likely to increase with system size. Therefore, we do not consider kinetic and potential energies separately for $N=1024$ where only DMC calculations were performed.  
All energies per electron are given in units of Ha. }
\label{dft_bcc}
\end{table}

\subsection{Liquid atomic and molecular hydrogen at high pressure}

We have used Coupled Electron-Ion Monte Carlo (CEIMC) to study high pressure hydrogen in the vicinity of the liquid-liquid, insulator-to-metal transition \cite{Carlotofinish,Hugo}. 
Here, the protonic configurations are sampled according to the classical Boltzmann distribution $\propto \exp[-\beta E_{BO}]$ where $E_{BO}$ is the Born-Oppenheimer energy of the protonic configuration determined from an electronic QMC estimate
\cite{CEIMC,RMP2012,entropy}. 
The nuclear configuration space was sampled using the VMC energy. The trial function consisted of a
Slater determinant of single electron orbitals for each spin component and a correlation part with single, two and three body Jastrows. The single electron orbitals are from self-consistent Kohn-Sham theory \cite{QE2009}, dressed by a backflow transformation. 
Analytical expressions from RPA for both correlation and backflow functions are employed~ \cite{bf,dewing1,ceimc,alder1} which exactly enforce the cusp conditions between all pairs of charges as well as the correct long-wavelength behavior of the charge oscillations. These are complemented by empirical expressions which preserve the correct short and long distance behavior and introduce few variational parameters that need to be optimized~\cite{dewing1,ceimc}. 

\begin{table}
\centering
\begin{tabular}{|c|c||c||c|c||c|c||c|}
\hline
$r_s$ & $N_e$ &$E_{TABC}$ & $\Delta T_{TABC}$  & $\Delta T_{BF}$ &$\Delta T$ & $\Delta V$ &$\Delta E$  \\
 \hline
 1.34 & 54 & -0.51591(7) & -0.000492 & -0.003912 &  0.000020 & 0.005250 & 0.005270 \\
  & 54-128 &                      &                    &                      &   -0.0005(10)   & 0.006(1)  & 0.0058(1) \\ \hline
  1.44 & 54 & -0.53679(8) & -0.000008 &-0.00363 & -0.000035 & 0.00444 & 0.00441 \\ 
      & 54-128 &   & & &0.003 (1) & 0.003(1) & 0.0056(2) \\ 
\hline
\end{tabular}
\caption{Finite size corrections for liquid atomic and molecular hydrogen at $T=1200K$. The kinetic energy corrections, $\Delta T=\Delta T_{TABC}+\Delta T_{BF}+\Delta T_U^{lr}$ sums up all contributions, whereas
the total potential energy corrections, are $\Delta V \equiv \Delta V_{lr}$. All size corrections correspond to energies per electron in Ha. }
\label{h_t1200}
\end{table}

In addition to the energy per particle, pressure corrections can be obtained from the kinetic and potential energy corrections using the virial theorem. The calculations were performed using TABC on a $4\times 4 \times 4$ grid for $N=54$ and $N=128$ hydrogen atoms. We have used
\eq{dTABC} to correct for finite size effects of the single particle kinetic energy imposing the DFT Fermi surface.
The potential energy can be written entirely in terms of the charged structure factor
\beq
S_c(\kvec)=\frac{1}{N_e} \left\langle \rho_{\kvec}^e \rho_{-\kvec}^e + \rho_\kvec^p \rho_{-\kvec}^p - 2 \rho_{\kvec}^e \rho_{-\kvec}^p \right\rangle
\eeq
averaged over electronic and protonic configurations,
and we have used the potential energy corrections, \eq{dVN}, with $\tilde{S}_N(\kvec)$ replaced by a
cubic spline interpolation of $S_c(\kvec)$.
The formula for the kinetic energy correction stemming from the Jastrow  cannot be simplified in terms of the charged structure 
factor only, but has been extended to include electron-electron and electron-proton components of the Jastrow, their correct
long-range are imposed in the VMC wavefunction
\footnote{Here, we consider classical protons. For quantum protons, proton-proton correlations must also be included in the kinetic energy size corrections.}.
The two-body backflow correction, Eq.~(\ref{TBF}), only involves the electron-electron structure factor, and $t$ has been obtained by calculating the kinetic energy of the backflow-free Slater determinant.

In table~\ref{h_t1200}, we illustrate size effects of the energy on two different systems. At the higher density, the system
is in the atomic phase with a metallic character, whereas the lower density is molecular and expected to be insulating.
Since the structure factor has a higher  peak for the molecular system around $k\approx 2.5$, the interpolation of it becomes less accurate and introduces a larger uncertainty in the size extrapolation than in the atomic liquid. 

\subsection{Hydrogen-Helium mixtures}

We now analyze some snapshot configurations of helium-hydrogen mixtures at high density, $r_s=1.10$ and $r_s=1.34$,
generated from an ab-initio quantum molecular dynamics simulations \cite{Ray}. For fixed nuclei positions, we have calculated the electronic energy using the QMCPACK \cite{qmcpack1,qmcpack2} simulation package based on a single Slater-Jastrow wavefunction
with single particle orbitals obtained from Quantum espresso \cite{QE2009} using the PBE functional (see Ref.~\cite{Ray,Raysup} for further details). 

Using a long-range optimized $u_k$ we can fit the long-range behavior of the electron-electron Jastrow $u_{ee}(k) =\alpha k^{-2}$ and the fluctuating electronic structure factor $\delta S_{ee}(k)=\beta k^2$ (using the pure estimator). We then
obtain the leading order corrections corresponding to the extrapolation of $k \to 0$ in \eq{TU}
\beq
\Delta T_U^{LO}=\lim_{k \to 0} \frac{\hbar^2 k^2 \rho_e u_{ee}^2(k) \delta S_{ee}(k)}{2 m_e V} =\frac{\hbar^2 \rho_e^2 \alpha^2 \beta}{2 m_e}\frac{1}{N}
\elabel{dTUab}
\eeq
and similar for the leading order potential energy corrections
\beq
\Delta V_{LO} = \lim_{k \to 0} \frac{v_k \delta S_{ee}(k)}{2 V} =  \frac{2 \pi \rho_e e^2 \beta}{N}
 \eeq
 In table \ref{h_he_rs1.10} we show the leading order size corrections and compare them to $\Delta V_{lr}$ obtained from interpolating only the pure estimator for the fluctuating structure factor as described in Sec.\ref{inhomo}.
The kinetic order corrections, $\Delta T_{lr}$, were obtained 
from the difference of the total energy corrections using the mixed estimator for the structure factor and the potential energy corrections from the pure estimator using a trial wavefunction without long-range components.
We compare our size effects with calculations of a $2 \times 2 \times 2$ supercell with frozen positions of the ions.
 
\begin{table}
\centering
\begin{tabular}{|c|c||c||c||c|c||c|c|c|}
\hline
$r_s$ & $x_{He}$ & $N_e$ & $\Delta T_{TABC}$  & $\Delta T_U^{LO}$ &$\Delta V_{LO}$ & $\Delta T_U^{lr}$ & $\Delta V_{lr}$&$\Delta E_{lr}$  \\
 \hline
1.10 & 6.7\% & 64 &-0.00059 & 0.00125 & 0.00470& 0.00143  & 0.00591 &0.00735 \\
        & & 64-512 & & & &0.0014(5) &0.0069(9)   &0.00816(3)\\ \hline
 & 21\% & 64 &-0.00024 &0.000273 &0.00452 &0.00130 & 0.00569&0.00700 \\
        & & 64-512 & && & -0.0009(10) &0.0095(10)  & 0.00859(2)\\      \hline \hline
 1.34 & 12\% & 64 &-0.00011 & 0.000178& 0.00343& 0.00103& 0.00426& 0.00530 \\
        & & 64-512 & & && 0.0017(8) &0.0047(8)  & 0.00632(3)\\      \hline
  & 21\% & 64 &-0.00008 & 0.000750& 0.00332& 0.00097& 0.00410&0.0051 \\
        & & 64-512 & & && 0.0018(7) &0.0046(7)  & 0.00640(2)\\      
\hline
\end{tabular}
\caption{Finite size corrections for snapshop configurations of a hydrogen-helium mixture at different densities, $r_s$, and different helium concentrations $x_{He}=N_{He}/(N_{He}+N_H)$. We give the size corrections based only on the $N_e=64$ electron system together with a linear extrapolation 
of the single particle ($\Delta T_{TABC}$) corrected energies using a $2\times 2 \times 2$ tiling of unit cell, denoted by $N_e=64-512$. All size corrections correspond to energies per electron in units of Ha.}
\label{h_he_rs1.10}
\end{table}

\section{Conclusions}

In this paper, we have discussed in detail the origin of finite size effects in quantum Monte Carlo calculations for electronic structure in extended systems. Based only on information on the wavefunction and the Hamiltonian, we have explicitly
shown the origin of size effects having either single particle or many-body character. We have proposed robust and
\textit{a priori}  methods
to reduce finite size errors without the need of performing calculations of different sizes nor relying on 
size extrapolations based on approximate calculations. 
Our methods do not assume any underlying symmetry of the unit cell  and have a comparable residual errors 
for solid and liquid structures.  Although we have used cubic supercells as
benchmark for our calculations, non-cubic systems can be treated as well.

Most of the explicit results are given for Slater-Jastrow wavefunctions, with or without backflow orbitals,
underlying most of the quantum Monte Carlo calculations for condensed matter. However, we want to stress that
our approach is more general and can be extended to many other situations not explicitly discussed,
e.g. Bose and Fermi systems at zero and finite temperature \cite{PIMC,finiteT}.

In particular, our method based on the interpolation of the static structure factor can be directly applied to any
system with pair-wise interaction to reduce the finite size error in the potential energy without 
any further assumption on the trial wavefunction
or the density matrix. More delicate are corrections of the two-body kinetic energy which require knowledge of the
long wavelength behavior of the
effective potentials of the underlying wavefunction, 
e.g. of the effective two-body Jastrow factor, $u_k$, or the effective backflow potential, $y_k$.
We also discussed and tested the method when such information was not explicitly available.

Our analysis of size corrections are also useful to judge the validity of different
size extrapolation schemes \cite{Rajagopal,MPC,Shiwei}. However,
as we have shown in the case of backflow wavefunctions, size-effects, in general, can depend on the form of the wavefunction,
an aspect usually not taken into account by heuristic approaches.
 Our theory of finite size extrapolation is based on reasonable assumptions of the correlation function which can be verified by additional calculations. Although our discussion was based on Quantum Monte Carlo calculations, 
our theory of finite size extrapolation should be applicable also to different
computational methods, e.g. FCIQMC \cite{Alavi}.

\section*{Appendix A: Finite size error in terms of analytic properties of Fourier transform}

Let us consider the general form of finite size error 
\beq
\Delta F = \left[ \int \frac{d\kvec}{(2\pi)^d} - \frac{1}{\Vol} \sum_\kvec \right] \tilde{f}_k
\eeq
where $\tilde{f}_k$ is obtained from the expression of the observable in Fourier space.
Assuming the existence of the Fourier transform $f(\rvec)$ of $\tilde{f}_k$, we have
\beq
\Delta F= \lim_{\rvec \to 0} \left[ f(\rvec) - \sum_{\nvec} f(\rvec + \nvec L) \right]
= \sum_{\nvec \ne 0} f(\nvec L)
\label{DF}
\eeq
where $\nvec$ is a d-dimensional vector with integer components. We see that for large systems,
the finite size error is directly connected 
to the long range behavior of $f$. 
 The functions $f$ and $\tilde{f}$ are usually isotropic for $r \to \infty$, and $k \to 0$, respectively, 
$f(\rvec) \sim \int dk k^{d-1} \int d\Omega_d  \tilde{f}_\kvec e^{i \kvec \cdot \rvec}$ and
$\tilde{f}_k \sim \int dr r^{d-1} \int d \Omega_d f(\rvec) e^{i \kvec \cdot \rvec}$, where $\Omega_d$ denotes the angular part of the volume integration.

From existence of the $n$th derivative of $\tilde{f}_k$ with respect to $k=|\kvec|$, we see that $|f(r)|$ must decay faster then $r^{-d-k+1}$ for $r \to \infty$,
and $\Delta F = {\cal O} \left(N^{-1-(k-1)/d} \right)$. If $\tilde{f}_k$ is an analytical function, it must be a regular function of $k^2$, its Fourier transform, $f(r)$, and therefore also $\Delta F$, decay exponentially with system size. Odd powers of $k$ at the origin indicate non-analytical behavior; 
we can reduce the finite size error by making the observable as smooth as possible, e.g. separating these non-analytical points from the integrand. In the case of long range (Coulomb) potentials, 
the original summations involved in energy or potential energy
exclude the term with $k=0$. However, since $\lim_{k \to 0} \tilde{f}_k$ remains in general finite, the inclusion of this term already
improves  the convergence to the thermodynamic limit. It can be further accelerated by separating out the non-analytic behavior around $k=0$ to make the reminder in the integrand more regular. The difficulty is to find a general way to split of part of $\tilde{f}_k$ around $k=0$ without introducing additional, artificial irregularities in the integrand.

For a classical Lennard-Jones potential, we can in general assume that $f$ decays at least as fast as the potential, $f(r)\sim v(r)\sim r^{-6}$
for large $r$, and Eq.(\ref{DF}) directly leads to a finite size error of order $N^{-2}$ in the potential energy for three dimensional systems.

For quantum systems, even in the case of short range potentials, phase fluctuations introduce long-range behavior in the correlation functions giving rise to non-analytical terms in potential and kinetic energy, independent of the statistics of the particles (Bosons or Fermions).  The sharp Fermi surface in a metallic state further introduces a discontinuity in the momentum distribution leading to slowly decaying oscillations in direct space, the origin of the shell effects in the kinetic energy.

\section*{Appendix B: The effective Jastrow potential and long-range structure factor}

Assume a quite general many-body wavefunction $\Psi(\Rvec)$,
$\Rvec \equiv (\rvec_1,\rvec_2,..\rvec_N)$. The mean value
of any observable $O(\Rvec)$ is given by
\beq
\langle O \rangle = \frac{\int d \Rvec O(\Rvec) |\Psi(\Rvec)|^2}{\int d \Rvec |\Psi(\Rvec)|^2}
\eeq
which we can re-write as
\beq
\langle O \rangle = 
\frac{
\int  d \Pi  O(\Pi)
\Psitilde^2(\Pi)}{\int d \Pi   \Psitilde^2(\Pi)}
\eeq
where $\Pi \equiv (\rho_{\kvec_1},\rho_{\kvec_2},...\rho_{\kvec_m})$, $\rho_\kvec\equiv \sum_{j=1}^N e^{i  \kvec \cdot \rvec_j}$, and 
\beq
\Psitilde^2(\Pi) \equiv \int d\Rvec  \prod_{n=1}^m \delta \! \! \left(\rho_{\kvec_n}-\sum_j e^{i \kvec_n \cdot \rvec_j} \right) 
|\Psi(\Rvec)|^2
\eeq
It is now natural to introduce an effective action $S_{\text{eff}} \equiv -\log |\Psitilde|$,  
with a functional form which respects the symmetry of the problem.
The simplest effective action
which respects translational invariance is given by
\beq
S_{\text{eff}} \simeq  \frac1{2\Vol} \sum_{\kvec \ne 0} \widetilde{u}_{\text{eff}}(\kvec) \rho_{\kvec} \rho_{-\kvec}
-\frac1{\Vol^2} \sum_{\kvec \qvec} w(\kvec,\qvec) \rho_{\kvec+\qvec} \rho_{-\kvec} \rho_{-\qvec}
\dots
\label{Seff}
\eeq
and reduces to
a simple Jastrow functional form in leading order.
Note that fermion effects are correctly included in this effective Jastrow factor.

Using the simplest effective action where we neglect the second term on the rhs of Eq.({\ref{Seff}) we get
\bea
S(k)&\equiv& \frac1N \langle \rho_{-\kvec} \rho_{\kvec} \rangle
\\
&\simeq&
\frac{\Vol}{2 N \widetilde{u}_{\text{eff}}(k)} =\frac1{2\rho \widetilde{u}_{\text{eff}}(k)}
\eea
which should be exact for $\kvec \to 0$ in the spirit of the RPA.

Applying this result to non-interacting Fermions described by a single Slater determinant, we see that the effective Jastrow
potential of the Slater determinant $[2\rho S_0(k)]^{-1}$ is completely determined by the corresponding 
non-interacting structure factor $S_0(k)$. For a general Slater-Jastrow wavefunction,
we therefore obtain \eq{skform} which relates the structure factor with the effective Jastrow potential in the long wavelength limit.

\section*{Appendix C: Derivation of backflow corrections}
Let us consider the  following backflow corrdinates in the orbitals of the determinant
\beq
\phi_{kj} \equiv \phi_\kvec(\qvec_j) \text{ with } \qvec_j \equiv \rvec_j + \etavec_j, \text{ and } \etavec_j =\frac{i}{\Vol} \sum_\kvec \kvec \, y_k \left[ e^{i \kvec \cdot \rvec_j} \rho_{-\kvec} - 1 \right]
\eeq
so that the laplacian of the determinant is
\beq
\nabla^2 D
=  \sum_{ij \alpha \beta} \frac{\partial^2 D}{\partial \qvec_i^\alpha \partial \qvec_j^\beta}  \left[ \nabla \qvec_i^\alpha \right] \cdot \left[ \nabla \qvec_j^\beta\right]
+ \sum_{i \alpha} \frac{\partial D}{\partial \qvec_i^\alpha} \nabla^2 \qvec_i^\alpha.
\eeq
Since $\phi_{kj}$ are (approximate) eigenfunctions of an effective Hamiltonian, terms with $i=j$ and $\alpha=\beta$ in the
first summation on the r.h.s.
are expected to dominate the expectation value
\beq
-\frac{\hbar^2}{2m_e} \left\langle \frac{1}{D} \nabla^2 D \right\rangle \approx \frac{t}{d} \sum_{i \alpha} \left\langle  \left[ \nabla \qvec_i^\alpha \right] \cdot \left[ \nabla \qvec_i^\alpha \right] \right\rangle
\eeq
where 
\beq
t=-\frac{\hbar^2}{2m_eN} \sum_{i \alpha}  \left\langle \frac{1}{D}  \frac{\partial^2}{\partial q_{i\alpha}^2} D\right\rangle
= \frac{1}{N} \sum_i  \left\langle \varepsilon_i - v_{eff}(\qvec_i)  \right\rangle
\eeq
is the single particle kinetic energy per particle using \eq{kineticE}.
We can now simplify
\beq
\nabla_n^\beta \qvec_i^\alpha
= \delta_{ni}
 \left[ \delta_{\alpha \beta} - \frac1\Vol \sum_\kvec k_\alpha k_\beta \, y_k \left( e^{i \kvec \cdot \rvec_n} \rho_{-\kvec} -1 \right) \right]
+ (1-\delta_{ni} ) \frac1\Vol \sum_\kvec k_\alpha k_\beta \, y_k e^{i \kvec \cdot( \rvec_i-\rvec_n)}
\eeq
and
\bea
\frac{1}{N}\sum_{i \alpha \beta}
\left\langle  \left[ \nabla \qvec_i^\alpha \right]^2 \right\rangle
&=& \frac1N \sum_{i\beta} \left\langle \left[ \delta_{\alpha \beta} - \frac1\Vol \sum_\kvec k_\alpha k_\beta \, y_k \left( e^{i \kvec \cdot \rvec_i} \rho_{-\kvec} -1 \right) \right]^2 
+ \sum_{n \beta}(1-\delta_{ni} ) \frac1{\Vol^2} \sum_{\kvec \kvec'}  k_\alpha k_\beta
k_\alpha' k_\beta'  \, y_k y_{k'}e^{i (\kvec+\kvec') \cdot( \rvec_i-\rvec_n)}
\right\rangle \nonumber \\
&=&
 \left[ 1- \frac{2}{d\Vol} \sum_\kvec k^2  \, y_k \left( S(k) -1 \right)  \right]
\nonumber
\\
&& + \frac{1}{d \Vol^2N} \sum_{\kvec \kvec' i}
 \left\langle   ( \kvec \cdot \kvec')^2 y_k y_{k'} \left( e^{i \kvec \cdot \rvec_i} \rho_{-\kvec}-1 \right)
  \left( e^{i \kvec' \cdot \rvec_i} \rho_{-\kvec'}-1 \right)
+ \sum_{n} (1-\delta_{ni} )   (\kvec \cdot \kvec')^2 \, y_k y_{k'}e^{i (\kvec+\kvec') \cdot( \rvec_i-\rvec_n)}
\right\rangle \nonumber \\
& \simeq &
 1- \frac{2}{d\Vol} \sum_\kvec k^2  \, y_k \left( S(k) -1 \right)  
 +\frac{\rho}{d \Vol} \sum_\kvec  k^4 y_k^2 \left(  S(k)+1 \right).
\eea

From this analysis, we can expect that backflow introduces another kinetic energy correction given by
\bea
\Delta T_{BF} & \approx & t \Delta s
\\
\Delta s& =& \left[ \int \frac{d \kvec}{(2\pi)^d} - \frac1\Vol \sum_\kvec \right]
\frac{k^2 y_k}{d} \left[ 2+\rho k^2 y_k - \left( 2 - \rho k^2 y_k  \right) S(k) \right]
\label{TBFA}
\eea
where $t$ is the mean kinetic energy of the single particle orbitals in the determinant.
From the long-range limit of the electron electron backflow \cite{bf,BFminus}, $y_q=- c(r_s)/nq^2$ with
$c(r_s) \approx 1+ 0.075 \sqrt{r_s}/(1+0.8\sqrt{r_s})$ for $d=3$,  frow which we  obtain the leading order term
\beq
\Delta s_{LO} = - \frac{1}{3N}
\eeq
or 
\beq
\Delta T_{BF}^{LO}= - \frac{t}{3N} 
 \eeq
 with $t \simeq 3 k_F^2/10m$ for a metal with spherical dispersion relation.

\section*{Appendix D: Extrapolation of potential energy based on Ewald summation and corrections due to non-analytic behavior of the structure factor}

To derive the long-range contributions for potential and kinetic energy, \eqtwo{dVN}{dTU},  
we have assumed that the long-range part of the underlying potentials can be separated  so that the
resulting expressions in Fourier space converge rapidly. In all examples provided in the main text, this separation was done
numerically using optimized potentials \cite{opt3d,opt2d}. 
Analytical expressions can be obtained for power-law potentials, in particular for the Coulomb $1/r$-interaction,
based on the method introduced by Ewald \cite{Ewald,Tildesley,ceperley78}.
In this expression, the periodic Coulomb potential inside a box of linear extension $L$, can be written as
\beq
v_{pp}(\rvec) =\frac{1}{\Vol} \sum_k v_k^{lr} e^{i \kvec \cdot \rvec} + \sum_{\nvec} v_{sr}(| \rvec + \nvec L|)
\eeq
with
\beq
v_k^{lr} = \frac{4 \pi}{k^2} e^{-k^2/4 \alpha^2}, \quad
v_{sr}(r) = \frac{\text{erfc}(\alpha r)}{r}
\eeq
where the parameter $\alpha$ controls the speed of convergence and $\nvec$ indicates the summation over all image charges in real space. In the following, we set $\alpha=\sqrt{k_c/L}$ and
cut-off the sum in reciprocal space at wave vector $k_c$ together with nearest-image convention in real space.

We illustrate the potential energy corrections for the homogeneous electron gas  within the Hartree-Fock approximation \cite{hf}.
Its structure factor in the thermodynamic limit is 
$S_{HF}(k)=3 k/4 k_F - k^3/16k_F^3$. Using GC-TABC the kinetic energy is exactly sampled. Additionally, the
finite size structure factor $S_N(k)$ is identical to the infinite one, $S_N(k)\equiv S_\infty(k)$,
on the discrete k-mesh compatible with the simulation box \cite{Lin01,fse}.
Thus, size effects are entirely due to the discretization error inside the calculation of the exchange energy. It is straighforward to calculate
the long-range contribution to the potential energy corrections. 

In deriving the energy correction formulas in the main text, we have assumed that the 
short range part of the pair correlation function is not modified by size corrections.
This would be the case if the
structure factor was an analytical function of $k$. However, within Hartree-Fock, 
the linear behavior of the structure factor strongly violates this assumption  in contrast to the expected $k^2$-behavior 
in more realistic calculations beyond Hartree-Fock. 
We can take into account this behavior by 
\beq
\Delta V_{sr} = \left[ \int \frac{d \kvec}{(2\pi)^d} - \frac{1}{\Vol} \sum_\kvec \right]v_k^{sr} \left[
\widetilde{S}_N(k) - \widetilde{S}_N(k_c) \right] \theta(k_c-k)
\elabel{Vsr}
\eeq
where $\widetilde{S}_N(k)$ is the interpolation of the structure factor imposing a vanishing derivative at the
cut-off $k_c$ and $v_k^{sr}=4 \pi (1-e^{-k^2/4 \alpha^2})/k^2 $ for the Ewald potential.
Notice, that we have inserted $\widetilde{S}_N(k_c)$ to force the integrand to vanish at $k_c$,
but the difference between the discrete sum and the integration vanishes to high precision by construction.

\begin{figure}
\includegraphics[width=0.5\textwidth]{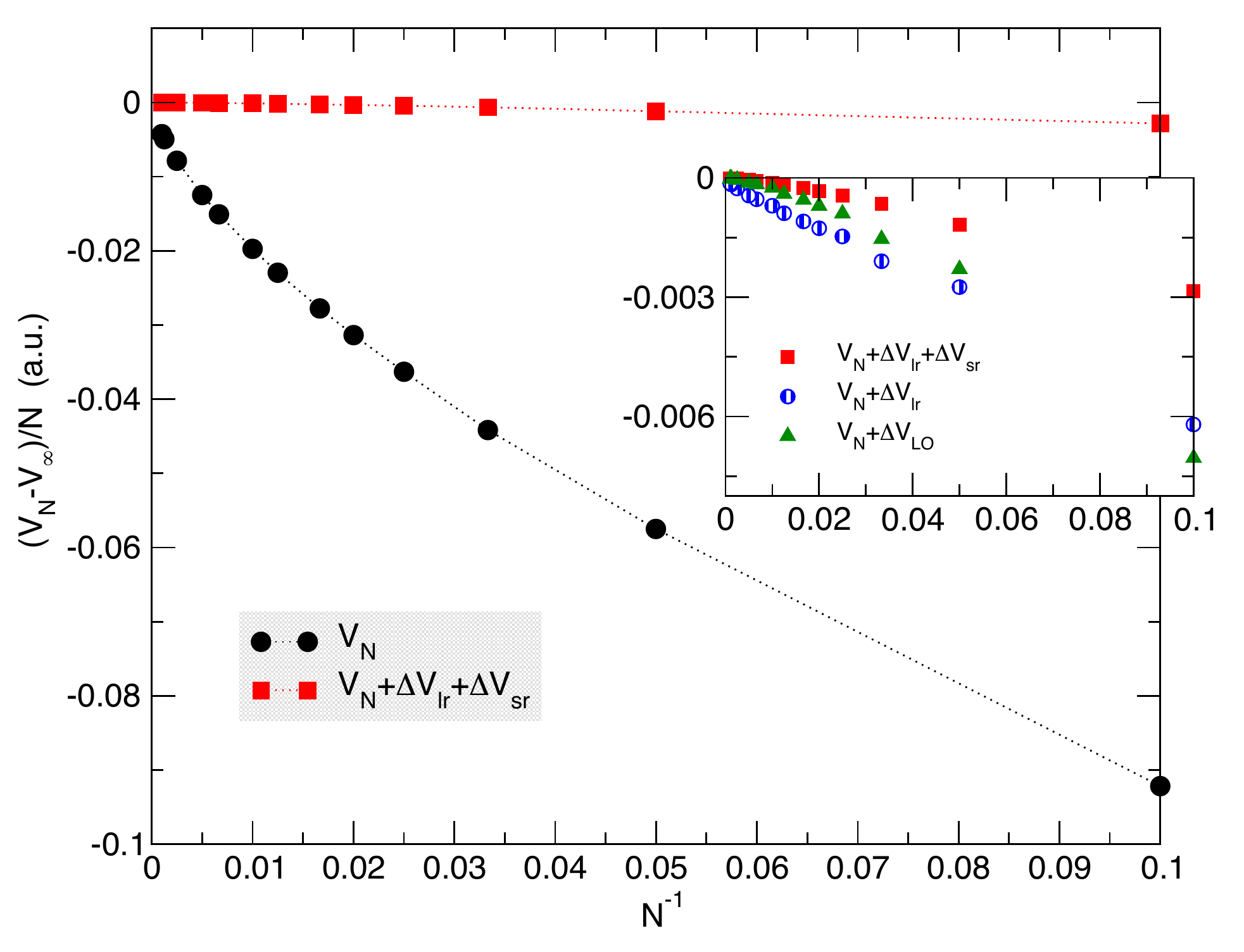}
  \caption{Differences of the potential energy per electron, $V_N$, in units of $Ha$, relative to the thermodynamic limit, $V_\infty$
  for the homogeneous electron gas at $r_s=1$ within the Hartree-Fock approximation. The black symbols are GCTABC results of the
  finite system with $N$ electrons, red symbols corresponds to the finite size corrected ones with long and short range corrections, $\Delta V_{lr}$ and $\Delta V_{sr}$, 
 using the Ewald method described in appendix D. In the inset we compare the long and short range potential corrections, $\Delta V_{lr}$ (pink symbols),  and $\Delta V_{sr}+ \Delta V_{lr}$ (red symbols) with the 
   leading order corrections (blue symbols) of Ref.~\cite{drummond}.
  Lines are guides to the eyes. For all size corrections, the value of the slope of the structure factor at the origin is estimated from its values at the discrete k-mesh,
  imposing the exact value, the energy difference can be further improved and
   $V_N+\Delta V_{lr}+\Delta_{sr}$ becomes equal to the thermodynamic limit on the scale of the figures.
}
\flabel{hf3D}
\end{figure}

In \fig{hf3D}, we show the finite size error in the exchange energy of the electron gas at $r_s=1$ within the Hartree-Fock approach
and the results of the potential energy corrections above in comparison with the leading order corrections \cite{drummond}.
For all size corrections, the value of the slope of the structure factor around the origin is estimated by finite difference of the finite size structure factor. Imposing the exact value of the slope, deviations to the thermodynamic limit of our best \textit{a priori} value
of the exchange energy, $V_N+\Delta V_{lr}+ \Delta V_{sr}$,  
are of order $\lesssim 10^{-5}$Ha for $N\gtrsim 10$.
Our procedure can therefore be considered as optimal.

In realistic calculations (beyond Hartree-Fock), screening effects strongly modify
long-range behavior of the structure factor compared to the Hartree-Fock behavior leading to $S(k) \sim k^2$ around $k=0$.
Although non-analytic behavior may still occur beyond leading order (terms of order $k^3$),  the corresponding size corrections $\Delta V_{sr}$ are expected to be much reduced. In practice, it is difficult to use \eq{Vsr} to correct for 
non-analytical terms beyond leading order, so that  we have only  taken into account $V_{lr}$ for size corrections in
3D. 
However, if we can extract the non-analytical behavior, $S(k)=a k^\alpha + b k^2 + \dots...$  around $k=0$, we can estimate
these corrections from the  asymptotic expansion
\beq
\Delta V_{sr} = a \left[ \int \frac{d \kvec}{(2\pi)^d} - \frac{1}{\Vol} \sum_\kvec \right] v_k^{sr} k^\alpha
\elabel{Vsralpha}
\eeq
These corrections are  of order $N^{-2}$ in 3D ($\alpha=3$) and $L^{-7/2} \sim N^{-7/4}$ in 2D ($\alpha=3/2$).

\section*{Acknowledgments}

CP was partially supported by the Italian Institute of Technology (IIT) under the SEED project Grant 259 SIMBEDD. 
MAM was supported by the U.S. Department of Energy at the Lawrence Livermore National Laboratory under Contract DE-AC52-07NA27344.  MAM, RC were supported through the Predictive Theory and Modeling for Materials and Chemical Science program by the Basic Energy Science (BES)
DMC and RC were supported by DOE grant  NA DE-NA0001789 and by the NanoSciences Fondation (Grenoble). MH and CP thank the Theory Group at ILL Grenoble for hospitality. Computer time was provided by PRACE projects 2011050781 and 2013091918 and by an allocation on the Blue Waters sustained-petascale computing project, supported by the National Science Foundation (award number OCI 07-25070) and the state of Illinois, and by CNRS-IDRIS, Project No. i2014051801.

\end{document}